\documentclass[a4paper,11pt]{article}
\pdfoutput=1

\usepackage{jcappub} 

\usepackage[T1]{fontenc}
\usepackage{comment}
\usepackage{xcolor}
\usepackage{arydshln}
\usepackage[normalem]{ulem}

\newcommand{\PaperI}{Paper~I}

\title{Probing Lorentz-violating electrodynamics with CMB polarization. II. Updated constraints and forecasts for the Simons Observatory and \textit{LiteBIRD}}

\author[a,*]{M. Lembo,}
\author[b,*]{L. Caloni\note[*]{Corresponding authors.},}
\author[c]{M. Gerbino,}
\author[d, e, f, g]{S. Giardiello,}
\author[h,i]{G. Gubitosi,}
\author[c]{M. Lattanzi,}
\author[l,c,m]{L. Pagano,}
\author[n,o]{N. Raffuzzi}

\affiliation[a]{Institut d’Astrophysique de Paris, UMR 7095, CNRS, and Sorbonne Universit\'e, 98 bis boulevard Arago, 75014 Paris, France}
\affiliation[b]{Faculdade de Ciências e Tecnologia and CFisUC, Departamento de Física,
Universidade de Coimbra, Rua Larga P-3004-516 Coimbra, Portugal}
\affiliation[c]{Istituto Nazionale di Fisica Nucleare (INFN), sezione di Ferrara, Via Giuseppe Saragat, 1, 44122, Ferrara}
\affiliation[d]{Department of Physics and Astronomy, School of Physical, Chemical and Environmental Sciences, Cardiff University, Cardiff CF24 3AA, UK}
\affiliation[e]{International School for Advanced Studies (SISSA), Via Bonomea 265, 34136 Trieste, Italy}
\affiliation[f]{Istituto Nazionale di Fisica Nucleare (INFN) Sezione di Trieste, Via Valerio 2, 34127 Trieste, Italy}
\affiliation[g]{Institute for Fundamental Physics of the Universe (IFPU), Via Beirut 2, 34151 Trieste, Italy}
\affiliation[h]{Dipartimento di Fisica Ettore Pancini, Universit\`a di Napoli “Federico II”, Complesso Univ. Monte S. Angelo, I-80126
Napoli, Italy}
\affiliation[i]{Istituto Nazionale di Fisica Nucleare, Sezione di Napoli, Complesso Univ. Monte S. Angelo, I-80126 Napoli, Italy}
\affiliation[l]{Dipartimento di Fisica e Scienze della Terra, Universita` di Ferrara, Via Giuseppe Saragat, 1, 44122, Ferrara}
\affiliation[m]{Institut d'Astrophysique Spatiale, CNRS, Univ. Paris-Sud, Universit\'{e} Paris-Saclay, B\^{a}t. 121, 91405 Orsay cedex, France}
\affiliation[n]{Physics Division, Lawrence Berkeley National Laboratory, Berkeley, CA 94720, USA}
\affiliation[o]{Department of Physics, University of California Berkeley, Berkeley, CA 94720, USA}

\emailAdd{margherita.lembo@cnrs.fr}
\emailAdd{luca.caloni@uc.pt}

\abstract{
We update the constraints on Lorentz-violating electrodynamics from the polarization of the cosmic microwave background (CMB). Within the minimal Standard Model Extension, we consider the renormalizable CPT-odd and CPT-even operators of the photon sector, which source cosmic birefringence and circular polarization, respectively.  Combining \textit{Planck}, Atacama Cosmology Telescope (ACT) and South Pole Telescope (SPT) temperature and $E$-mode data with $B$-mode measurements from BICEP/\textit{Keck}, SPT-3G D1 and SPTpol, we constrain the CPT-odd coefficients to $|k^{(3)}_{(V)00}|<7.75\times10^{-45}$\,GeV and $|\mathbf{k_{AF}}|<3.82\times10^{-45}$\,GeV at 95\% credible level. These are the strongest bounds to date from any probe, and are driven by the SPT-3G D1 $B$-mode data. 
For the first time in this type of analysis, we also include the \textit{Planck} PR4 $EB$ spectrum, which yields a mild preference for a non-zero $|k^{(3)}_{(V)00}|$.
For the CPT-even coefficient, we find $k_{F,E+B}<5.99\times10^{-32}$. Including the latest circular polarization measurements from the Cosmology Large Angular Scale Surveyor (CLASS) tightens this bound by about 7\%. This is the first time that CMB circular polarization data contribute to constraining this coefficient. Finally, we forecast the sensitivity of the Simons Observatory (SO) and \textit{LiteBIRD}. With linear polarization only, the two experiments are complementary, with SO tightening the CPT-even bound by about 30\% and \textit{LiteBIRD} the CPT-odd ones by 40 to 45\%. Exploiting the sensitivity to circular polarization provided by the non-idealities of its half-wave plate, \textit{LiteBIRD} could also improve the CPT-even bound by more than an order of magnitude.
}

\begin{document}
\maketitle
\flushbottom

\section{Introduction}
\label{sec:intro}

Lorentz invariance is one of the fundamental symmetries of modern physics, underlying both the Standard Model of particle physics and general relativity. Several approaches to quantum gravity nonetheless suggest that this symmetry might be violated at very high energies, with residual violations potentially surviving down to observable scales~\cite{Mattingly:2005re,Amelino-Camelia:2008aez,Addazi:2021xuf}. Searching for such violations is therefore a powerful way to probe physics beyond the Standard Model.

A general and widely used framework for parametrizing possible deviations from exact Lorentz invariance is the Standard Model Extension (SME)~\cite{Colladay:1998fq,Kostelecky:2002hh}, an effective field theory that extends the Standard Model by adding all Lorentz-violating (LV) operators consistent with the underlying gauge symmetries, organized by their mass dimension. In the photon sector, the leading renormalizable operators are a CPT-odd, dimension-3 term and a CPT-even, dimension-4 term. 
Both modify the propagation of photons, with an effect that grows with the distance they travel, making astrophysical and cosmological sources particularly sensitive probes.
Among these, the cosmic microwave background (CMB) is the most distant source of electromagnetic radiation available in nature and is linearly polarized by Thomson scattering at recombination and reionization~\cite{Hu:2001bc}. Any modification of its polarization induced by these operators is therefore integrated over the longest possible path, so that even tiny Lorentz violations can leave a measurable imprint~\cite{Kostelecky:2007zz, Caloni:2022kwp}.
Concretely, the CPT-odd operator induces a rotation of the linear polarization plane of CMB photons, an effect known as cosmic birefringence; its time component sources an isotropic rotation, while its space components source an anisotropic one. The isotropic rotation gives rise to parity-violating $EB$ and $TB$ correlations, absent in the standard cosmological model, while both induce a leakage of $E$- into $B$-mode polarization. Isotropic birefringence is particularly relevant today, since analyses of \textit{Planck} and Wilkinson Microwave Anisotropy Probe (WMAP) data have found hints of a non-zero rotation angle~\cite{Minami:2020odp,Eskilt:2022cff,Diego-Palazuelos:2022dsq}, recently supported by a consistent measurement from the sixth data release (DR6) of the Atacama Cosmology Telescope (ACT)~\cite{Diego-Palazuelos:2025dmh}. If confirmed to be of cosmological origin, and not due to a residual miscalibration of the detector polarization angles, such a signal would be evidence of parity-violating physics beyond the Standard Model. Proposed mechanisms include a pseudoscalar coupling between photons and an axion-like field~\cite{Carroll:1989vb,Carroll:1998zi,Li:2008tma,Pospelov:2008gg}, which may constitute dark energy or dark matter~\cite{Fujita:2020ecn,Obata:2021nql,Gasparotto:2022uqo,Murai:2022zur,Gasparotto:2023psh} or form networks of domain walls and cosmic strings~\cite{Takahashi:2020tqv,Kitajima:2022jzz,Jain:2021shf,Ferreira:2023jbu}, a vector Chern-Simons coupling between photons and the neutrino current in the presence of a large lepton asymmetry~\cite{Geng:2007va,Zhou:2023aqz,Kasai:2026qty}, and the LV electrodynamics considered in this work.
The CPT-even operator, instead, converts linear into circular polarization, sourcing a non-zero $VV$ spectrum. Circular polarization is not generated by scattering at recombination, and the mechanisms that can produce it within the standard cosmological model yield a negligible signal~\cite{Montero-Camacho:2018vgs, King:2016exc}. 
A detection of $V$ modes in the CMB would therefore be a signature of beyond-$\Lambda$CDM physics~\cite{Alexander:2008fp,Ejlli:2016avx,Tizchang:2016vef,Ejlli:2017uli,Inomata:2018vbu,Bartolo:2019eac,Hoseinpour:2020hic,Bavarsad:2009hm,Kushwaha:2025yja}, including Lorentz violation.

In~\cite{Caloni:2022kwp} (hereafter Paper~I), we set up the framework to test LV electrodynamics with CMB polarization. Building on the effective susceptibility tensor formalism developed in~\cite{Lembo:2020ufn}, we mapped the LV operators onto phenomenological parameters that source isotropic and anisotropic birefringence and circular polarization in the CMB, and constrained them with data from \textit{Planck} 2018~\cite{Planck:2018vyg}, BICEP/\textit{Keck} 2018~\cite{BICEP:2021xfz} and ACT DR4~\cite{ACT:2020gnv}. In this work, we update and extend the analysis of \PaperI\ along several lines. First, we replace ACT DR4 with DR6~\cite{AtacamaCosmologyTelescope:2025blo}, and include data from the South Pole Telescope (SPT). From the third-generation receiver SPT-3G, we use the recent small-scale temperature and polarization measurements of the D1 data release~\cite{SPT-3G:2025bzu} and the dedicated $B$-mode measurements from the same dataset~\cite{SPT-3G:2025vtb}. We complement these with the earlier $B$-mode measurements from the previous-generation receiver SPTpol~\cite{SPT:2019nip}. Second, for the first time in this type of analysis, we include the parity-violating $EB$ spectrum from the \textit{Planck}~PR4 \texttt{Lollipop} likelihood~\cite{Tristram:2020wbi, Tristram:2021tvh}, which is expected to vanish in the standard cosmological model and is therefore a particularly clean probe of isotropic birefringence. Third, we use the latest circular polarization measurements from the Cosmology Large Angular Scale Surveyor (CLASS) at 90, 150 and 220~GHz~\cite{Essinger-Hileman:2026tgi}. Finally, we forecast the sensitivity of the ground-based Simons Observatory (SO)~\cite{SimonsObservatory:2018koc}, which has already started science observations, and of the \textit{LiteBIRD} satellite~\cite{LiteBIRD:2022cnt}. The two experiments probe complementary angular scales. SO measures temperature and polarization anisotropies from degree to arcminute scales, while \textit{LiteBIRD} will target the largest angular scales over the full sky. Following the approach we introduced in~\cite{Raffuzzi:2024wyh}, we show that the non-idealities of the \textit{LiteBIRD} half-wave plate could provide competitive sensitivity to circular polarization, and hence to the CPT-even operator.

The paper is organized as follows. In section~\ref{sec:theory} we summarize the LV model, its impact on the CMB power spectra and the mapping between the phenomenological parameters and the SME coefficients. In section~\ref{sec:methodology} we describe the datasets and the methodology adopted for current data and forecasts, and in section~\ref{sec:results} we present the resulting constraints and recast them in terms of the SME coefficients. We conclude in section~\ref{sec:conclusion}. Appendices~\ref{app:B-modes} and~\ref{app:calibration} discuss the role of individual $B$-mode datasets and the degeneracy with the polarization calibration parameters of ACT and SPT.


\section{Model and CMB spectra}
\label{sec:theory}
In this section we summarize the main aspects of the Lorentz-violating model relevant for our analysis and its impact on the CMB spectra, referring the reader to~\cite{Caloni:2022kwp, Lembo:2020ufn} for more details.
We work within the minimal SME~\cite{Colladay:1998fq,Kostelecky:2002hh}, considering renormalizable operators that extend Maxwell electrodynamics. The corresponding action reads
\begin{equation}
	\label{eq:action-LV}
	\mathcal{S} = \int d^4 x \sqrt{-g}\left[ -\frac{1}{4} F_{\mu\nu}F^{\mu\nu} + \frac{1}{2}\varepsilon^{\alpha\beta\mu\nu}A_\beta (k_{AF})_\alpha F_{\mu\nu} - \frac{1}{4} (k_F)^{\alpha\beta\mu\nu} F_{\alpha\beta}F_{\mu\nu}\right] \, ,
\end{equation}
where the vector $(k_{AF})_\alpha$ and the tensor $(k_F)^{\alpha\beta\mu\nu}$ encode the CPT-odd and CPT-even Lorentz-violating contributions, respectively. The corresponding operators have mass dimension $d=3$ and $d=4$, respectively, so that $k_{AF}$ has dimensions of mass while $k_F$ is dimensionless. As shown in \PaperI, these operators modify the propagation of CMB photons between the last scattering surface and today, leaving distinct imprints on the observed polarization. 

\begin{figure}[t!]
    \centering
    \includegraphics[width=0.49\linewidth]{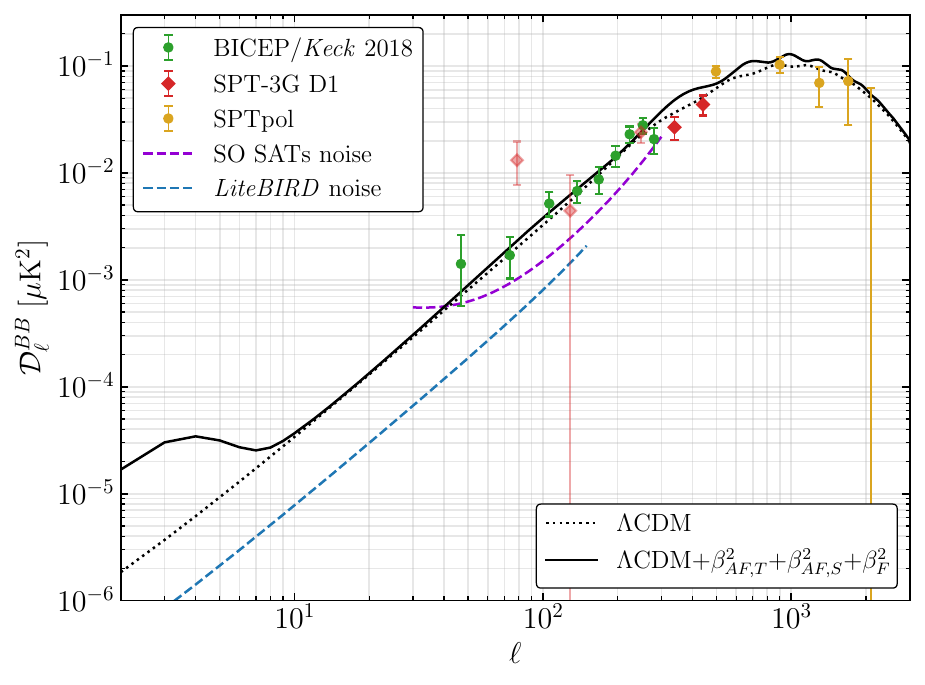}
    \includegraphics[width=0.49\linewidth]{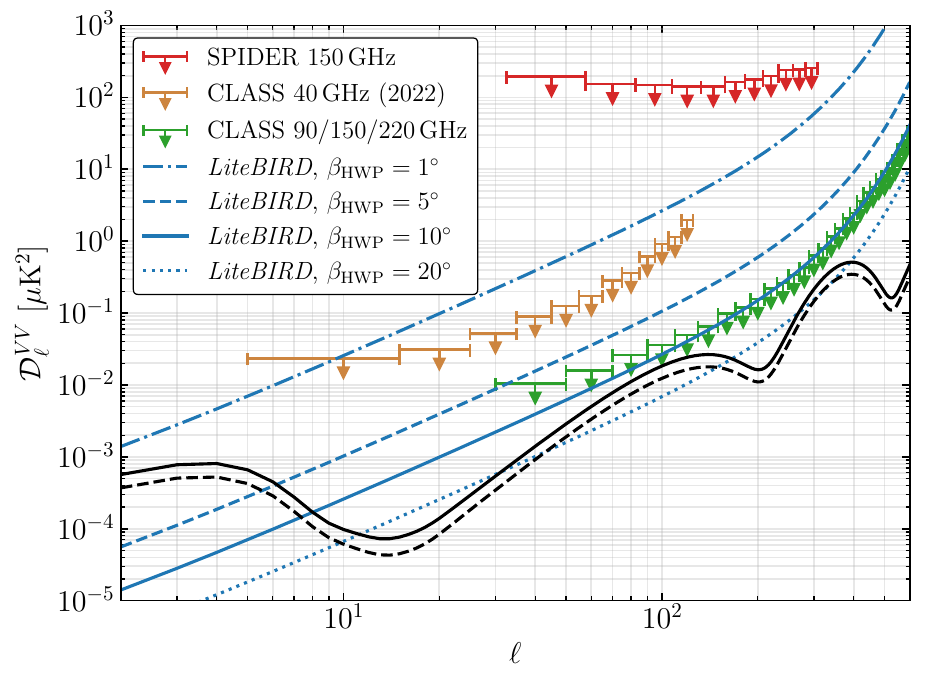}
    \caption{\emph{Left:} CMB $B$-mode power spectrum for the $\Lambda$CDM model (lensing only) and for models with Lorentz-violating parameters $\beta^2_{AF,T}$, $\beta^2_{AF,S}$, and $\beta^2_{F}$ each fixed to their 95\% upper limit obtained from the SPA+BK18+SPT$BB$+SPTpol$BB$ dataset combination (see section~\ref{sec:current_data} for the definition of the dataset labels). Data points show $B$-mode measurements from BK18~\cite{BICEP:2021xfz}, SPT-3G D1~\cite{SPT-3G:2025vtb} and SPTpol~\cite{SPT:2019nip}. SPTpol data points are restricted to $\ell > 450$ to avoid overlap with SPT-3G D1. Semi-transparent red diamonds show SPT-3G D1 data points in the multipole range covered by BK18; these are not used when BK18 is included in the analysis (see section~\ref{sec:current_data} for more details). We also show the SO SATs and \textit{LiteBIRD} noise curves, obtained by combining the frequency channels with inverse-noise weighting, as described in section~\ref{sec:forecasts}. In particular, for SO SATs we adopt the baseline sensitivity and pessimistic $1/f$ noise as described in~\cite{SimonsObservatory:2018koc}.
    \emph{Right:} CMB $V$-mode power spectrum for models with LV parameters fixed to their 95\% upper limits obtained from current temperature and linear polarization data (SPA+BK18+SPT$BB$+SPTpol$BB$, black solid) and from the \textit{LiteBIRD} $TEB$-only forecast (black dashed). Data points show the 95\% CL upper limits on the $VV$ spectrum from the combination of the CLASS 90, 150 and 220~GHz measurements used in this work~\cite{Essinger-Hileman:2026tgi}, compared with previous limits from CLASS at 40~GHz~\cite{Eimer:2023esh} and from SPIDER at 150~GHz~\cite{SPIDER:2017kwu}. The noise curves in blue show the expected \textit{LiteBIRD} sensitivity to $V$ modes for different values of the HWP phase shift, $\beta_{\rm HWP} = 1^\circ$, $5^\circ$, $10^\circ$, and $20^\circ$ (see section~\ref{sec:forecasts} for more details).}
    \label{fig:CMB_spectra}
\end{figure}

The impact of these operators on the CMB power spectra is encoded in four phenomenological parameters, $\beta^2_{AF,T}$, $\beta^2_{AF,S}$, $\beta^2_{F,E}$ and $\beta^2_{F,B}$; the first two are built from the time and space components of $k_{AF}$, respectively, while the latter two are built from specific contractions of $k_F$ (see section~\ref{sec:connections-beta-k} and \PaperI\ for more details). In terms of these parameters, the observed CMB spectra, $C_\ell^{XY}$ with $X,Y\in\{T,E,B,V\}$, are related to the spectra in the absence of Lorentz violation, $\widetilde{C}_\ell^{XY}$, by~\cite{Caloni:2022kwp}
\begin{align}
    \label{eq:clTE}
    C^{TE}_\ell &= \left(1-\frac{\mathcal{Z}}{2}\right)\widetilde{C}^{TE}_{\ell} \,, \\
    \label{eq:EE}
    C^{EE}_\ell &= \left(1-\mathcal{Z}\right)\widetilde{C}^{EE}_{\ell}
    + \sum_{\ell_1} \mathcal{K}^{11}_{\ell_1\ell}\widetilde{C}^{EE}_{\ell_1}
    + \sum_{\ell_1} \mathcal{K}^{22}_{\ell_1\ell}\widetilde{C}^{BB}_{\ell_1} \,, \\
    \label{eq:BB}
    C^{BB}_\ell &= \left(1-\mathcal{Z}\right)\widetilde{C}^{BB}_{\ell}
    + \sum_{\ell_1} \mathcal{K}^{11}_{\ell_1\ell}\widetilde{C}^{BB}_{\ell_1}
    + \sum_{\ell_1} \mathcal{K}^{22}_{\ell_1\ell}\widetilde{C}^{EE}_{\ell_1} \,, \\
    \label{eq:EB}
    C^{EB}_\ell &= \sqrt{\beta^2_{AF,T}}\left(\widetilde{C}^{EE}_{\ell}-\widetilde{C}^{BB}_{\ell}\right) \,, \\[0.3cm]
    \label{eq:TB}
    C^{TB}_\ell &= \sqrt{\beta^2_{AF,T}}\,\widetilde{C}^{TE}_{\ell} \,, \\[0.3cm]
    \label{eq:clVV}
    C_\ell^{VV} &= \sum_{\ell_1} \mathcal{K}^{33}_{\ell_1\ell}\widetilde{C}^{EE}_{\ell_1}
    + \sum_{\ell_1} \mathcal{K}^{44}_{\ell_1\ell}\widetilde{C}^{BB}_{\ell_1} \,, \\
    C_\ell^{EV} &= C_\ell^{BV} = 0 \, ,
\end{align}
where we have defined
\begin{align}
    \label{eq:Z}
    \mathcal{Z} &= \beta^2_{AF,T}+\beta^2_{AF,S} + \frac{\beta^2_{F,E}+\beta^2_{F,B}}{4} \,, \\
    \label{eq:K11}
    \sum_{\ell_1} \mathcal{K}^{11}_{\ell_1\ell}\widetilde{C}^{XX}_{\ell_1} &= \beta^2_{AF,S}\,\frac{4}{\ell+\ell^2}\,\widetilde{C}^{XX}_{\ell} \,, \\
    \label{eq:K22}
    \sum_{\ell_1} \mathcal{K}^{22}_{\ell_1\ell}\widetilde{C}^{XX}_{\ell_1} &= \beta^2_{AF,T}\,\widetilde{C}^{XX}_{\ell} + \beta^2_{AF,S}\left[ \frac{\ell^2-4}{\ell(2\ell+1)}\widetilde{C}^{XX}_{\ell-1} + \frac{(\ell-1)(\ell+3)}{(\ell+1)(2\ell+1)}\widetilde{C}^{XX}_{\ell+1} \right] \,, \\
    \label{eq:K33}
    \sum_{\ell_1} \mathcal{K}^{33(44)}_{\ell_1\ell}\widetilde{C}^{XX}_{\ell_1}
    &= \beta^2_{F,B(E)}\left[ \frac{(\ell-2)(\ell-3)}{4(4\ell^2-1)}\widetilde{C}^{XX}_{\ell-2} + \frac{3(\ell^2+\ell-2)}{2(4\ell^2+4\ell-3)}\widetilde{C}^{XX}_{\ell} \right. \\
    &\left.\qquad + \frac{(\ell+3)(\ell+4)}{4(4\ell^2+8\ell+3)}\widetilde{C}^{XX}_{\ell+2} \right]
    + \beta^2_{F,E(B)}\left[ \frac{\ell-2}{2(2\ell+1)}\widetilde{C}^{XX}_{\ell-1} + \frac{\ell+3}{2(2\ell+1)}\widetilde{C}^{XX}_{\ell+1}\right] \, . \nonumber
\end{align}

By inspecting eqs.~\eqref{eq:clTE}--\eqref{eq:K33}, we can identify the effect of each class of operators. The CPT-odd parameters $\beta^2_{AF,T}$ and $\beta^2_{AF,S}$ source the well-known cosmic birefringence effect, inducing a mixing between the $EE$ and $BB$ spectra. In particular, $\beta^2_{AF,T}$ produces isotropic birefringence, which also gives rise to non-vanishing $EB$ and $TB$ spectra (eqs.~\eqref{eq:EB}--\eqref{eq:TB}), while $\beta^2_{AF,S}$ generates anisotropic birefringence, coupling each multipole $\ell$ to its neighbours $\ell\pm1$. The effect of the leakage of $E$ into $B$ modes is shown in the left panel of figure~\ref{fig:CMB_spectra} as two distinct features: one at $\ell \gtrsim 300$, accessible to ground-based $B$-mode measurements at intermediate scales, and an additional excess at very large angular scales, present even in the absence of primordial tensor modes, which is potentially relevant for a future satellite experiment such as \textit{LiteBIRD}. The CPT-even parameters $\beta^2_{F,E}$ and $\beta^2_{F,B}$ instead source a non-zero $VV$ spectrum from linear polarization spectra, coupling multipoles between $\ell-2$ and $\ell+2$; notably, the $VV$ spectrum is, in principle, the only one sensitive to $\beta^2_{F,E}$ and $\beta^2_{F,B}$ individually, since all the other spectra only depend on their sum (see section~\ref{sec:VV_degeneracy}). The resulting $VV$ spectrum is shown in the right panel of figure~\ref{fig:CMB_spectra}, for values of the LV parameters corresponding to their 95\% credible level upper limits from current temperature and linear-polarization data and from the \textit{LiteBIRD} forecast.
Finally, all the parameters rescale the $TE$, $EE$ and $BB$ spectra through the common factor $\mathcal{Z}$ defined in eq.~\eqref{eq:Z}. Through this rescaling, linear-polarization measurements also acquire sensitivity to the CPT-even parameters.

The spectra presented in this section, and used throughout the rest of the analysis, are computed with \texttt{CAMB\_LV}, a modified version of \texttt{CAMB}~\cite{Lewis:1999bs,Howlett:2012mh} that extends the \texttt{CAMB\_CPT} code used in \PaperI.\footnote{With respect to \texttt{CAMB\_CPT}, \texttt{CAMB\_LV} includes the computation of the $EB$, $TB$ and $VV$ spectra. The code is publicly available at \url{https://github.com/mlembo00/CAMB_LV}.}

\subsection[On the degeneracy between ${\beta^2_{F,E}}$ and ${\beta^2_{F,B}}$]{On the degeneracy between $\boldsymbol{\beta^2_{F,E}}$ and $\boldsymbol{\beta^2_{F,B}}$}
\label{sec:VV_degeneracy}

Although $V$-mode polarization is, in principle, the only observable sensitive to $\beta^2_{F,E}$ and $\beta^2_{F,B}$ individually, we demonstrate that even for an ideal, cosmic-variance-limited experiment the $VV$ power spectrum cannot separate them in practice, and only their sum is a well-constrained quantity.

This can be understood from the structure of eq.~\eqref{eq:clVV}. Defining the 
two multipole-mixing kernels
\begin{align}
    \mathcal{A}[X]_\ell &\equiv \frac{(\ell-2)(\ell-3)}{4(4\ell^2-1)}X_{\ell-2} + \frac{3(\ell^2+\ell-2)}{2(4\ell^2+4\ell-3)}X_{\ell} + \frac{(\ell+3)(\ell+4)}{4(4\ell^2+8\ell+3)}X_{\ell+2} \, , \\
    \mathcal{B}[X]_\ell &\equiv \frac{\ell-2}{2(2\ell+1)}X_{\ell-1} + \frac{\ell+3}{2(2\ell+1)}X_{\ell+1} \, ,
\end{align}
eq.~\eqref{eq:clVV} can be recast, in terms of
\begin{equation}
    \label{eq:beta-F}
    \beta^2_F\equiv\beta^2_{F,E}+\beta^2_{F,B} \,  \quad\text{and}\quad  \Delta\beta^2_F\equiv\beta^2_{F,E}-\beta^2_{F,B},
\end{equation}
as
\begin{equation}
    \label{eq:clVV-SD}
    C_\ell^{VV} = \frac{\beta^2_F}{2}\big(\mathcal{A}+\mathcal{B}\big)\big[\widetilde C^{EE}+\widetilde C^{BB}\big]_\ell 
    + \frac{\Delta\beta^2_F}{2}\big(\mathcal{B}-\mathcal{A}\big)\big[\widetilde C^{EE}-\widetilde C^{BB}\big]_\ell \, ,
\end{equation}
where $\mathcal{A}+\mathcal{B}$ acts on $\widetilde C^{EE}+\widetilde C^{BB}$, and $\mathcal{B}-\mathcal{A}$ on $\widetilde C^{EE}-\widetilde C^{BB}$.
Since $\widetilde C^{BB}_\ell \ll \widetilde C^{EE}_\ell$, both terms in eq.~\eqref{eq:clVV-SD} are dominated by essentially the same spectrum, $\widetilde C^{EE}_\ell$. Moreover, $\mathcal{A}$ and $\mathcal{B}$ both mix each multipole with nearby ones, and on a smooth spectrum such as $\widetilde C^{EE}_\ell$ they give nearly the same result. The prefactor of $\Delta\beta^2_F$, $\mathcal{B}-\mathcal{A}$, therefore nearly vanishes, while that of $\beta^2_F$, $\mathcal{A}+\mathcal{B}$, does not. The two operators differ appreciably only at the lowest multipoles, $\ell\lesssim10$, which are also the scales where cosmic variance is largest. This explains why $\beta^2_F$ is well constrained while $\Delta\beta^2_F$ is not.

To quantify this, we compare the two extreme cases at fixed $\beta^2_F$, in which only one of the two parameters is non-zero, namely either $\beta^2_{F}=\beta^2_{F,E}$ or $\beta^2_F=\beta^2_{F,B}$. 

\begin{figure}[!t]
    \centering
    \includegraphics[width=0.7\linewidth]{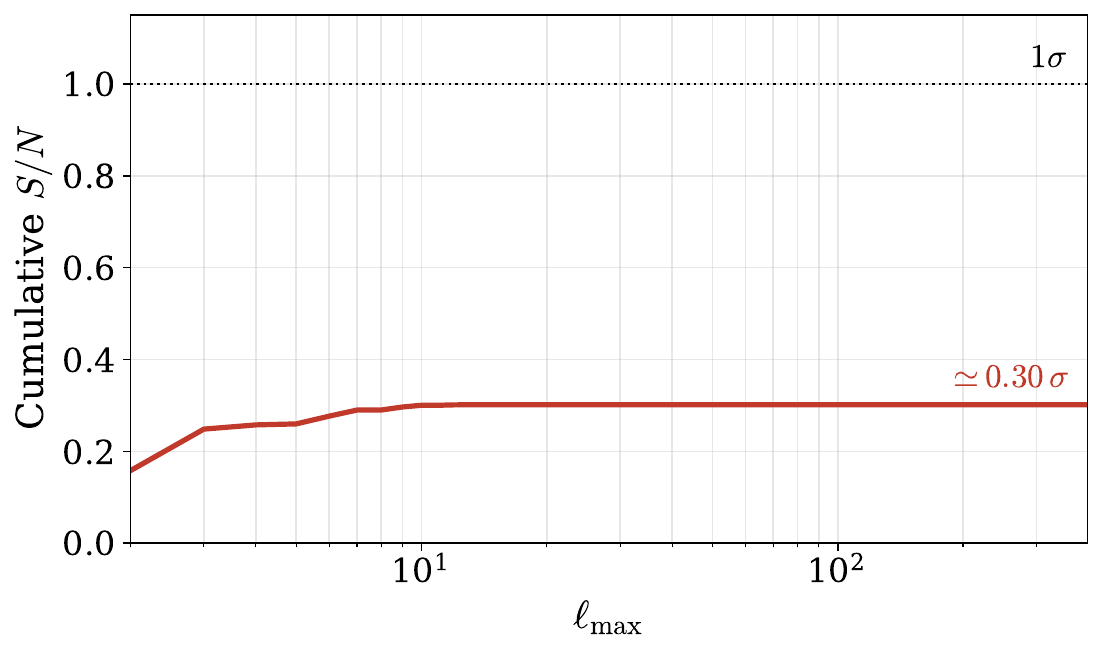}
    \caption{Cumulative signal-to-noise ratio, as a function of the maximum multipole $\ell_{\max}$, as defined in eq.~\eqref{eq:signal_to_noise}. This is shown for a cosmic-variance limited experiment with $f_{\rm sky}=1.0$. The dotted line marks the $1\sigma$ level.}
    \label{fig:VV_SN}
\end{figure}

For these two cases, we compute the cumulative signal-to-noise ratio
\begin{equation}
    \label{eq:signal_to_noise}
    S/N(\ell_{\max}) = \Bigg[\sum_{\ell=2}^{\ell_{\max}}
    \left(\frac{\Delta C^{VV}_\ell}{\sigma_\ell}\right)^2\Bigg]^{1/2},
\end{equation}
where $\Delta C^{VV}_\ell$ is the difference between the two $VV$ spectra and $\sigma_\ell=\sqrt{2/[(2\ell+1)f_{\rm sky}]}\,C^{VV}_\ell$ is the cosmic-variance uncertainty of the spectrum with $\beta^2_{F,E}=\beta^2_{F,B}$. We show it in figure~\ref{fig:VV_SN} for $f_{\rm sky}=1.0$. Since every $VV$ spectrum at fixed composition is proportional to $\beta^2_F$, $\beta^2_F$ cancels in this ratio and the result does not depend on the amplitude of the signal.
The signal-to-noise ratio reaches only $\simeq0.3$ at $\ell_{\max}\simeq10$ and does not increase further at higher multipoles. Even an ideal experiment would therefore be unable to distinguish the two extreme cases using the $VV$ spectrum, which only captures the diagonal part of the $V$-mode covariance. Additional information could in principle come from the correlations between different multipoles induced by the anisotropic components of $k_F$.

For this reason, we retain only the term proportional to $\beta^2_F$ in eq.~\eqref{eq:clVV-SD}, i.e. we set $\Delta\beta^2_F=0$, and report constraints on $\beta^2_F$ throughout this work.

\subsection[Mapping ${\beta^2_{AF,T}}$, ${\beta^2_{AF,S}}$ and ${\beta^2_F}$ to the SME coefficients]{Mapping $\boldsymbol{\beta^2_{AF,T}}$, $\boldsymbol{\beta^2_{AF,S}}$ and $\boldsymbol{\beta^2_F}$ to the SME coefficients}
\label{sec:connections-beta-k}
In this section we provide the explicit mapping between the phenomenological $\beta$ parameters and the physical coefficients appearing in the action~\eqref{eq:action-LV}; we refer the reader to \PaperI\ for the full derivation. The time and space components of $k_{AF}$, usually recast as 
$k^{(3)}_{(V)00}\equiv-\sqrt{4\pi}(\bar k_{AF})^0$ and $|\mathbf{k_{AF}}|$, are related to 
$\beta^2_{AF,T}$ and $\beta^2_{AF,S}$ by
\begin{align}
    \label{eq:kv00}
    |k^{(3)}_{(V)00}| &= \frac{\sqrt{\pi}}{2c(\tau_0-\tau_{\rm LS})}\sqrt{\beta^2_{AF,T}} \simeq 4.09\times10^{-43}\sqrt{\beta^2_{AF,T}}\ \text{GeV} \, , \\
    \label{eq:kaf}
    |\mathbf{k_{AF}}| &\simeq 2.00\times10^{-43}\sqrt{\beta^2_{AF,S}}\ \text{GeV} \, ,
\end{align}
where the bar denotes the line-of-sight average and $c(\tau_0-\tau_{\rm LS}) \simeq 13871$~Mpc is computed for the mean $\Lambda$CDM parameters of the SPA+BK18+SPT$BB$+SPTpol$BB$ combination (see section~\ref{sec:current_data} for the definition of the dataset labels) with the LV parameters fixed to zero.\footnote{Adopting the \textit{Planck} 2018 cosmology, the value of $c(\tau_0-\tau_{\rm LS})$ changes by a negligible amount ($\lesssim 0.03 \%$). We therefore always adopt the value obtained for the baseline combination.} The relation for the CPT-even coefficient is instead frequency dependent\footnote{While $k_{AF}$, which has dimensions of mass, induces a phase shift proportional to $k_{AF}L$ along a path of length $L$, the dimensionless $k_F$ induces one proportional to $k_F\,\nu L$, which grows with the photon frequency.}:
\begin{equation}
    \label{eq:kFEB}
    k_{F,E+B} \equiv \left(2\bar k_{F,E}^2+\frac{\bar k_{F,B}^2}{3}\right)^{1/2} \simeq 4.38\times 10^{-29}\left(\frac{\nu}{\rm GHz}\right)^{-1}\sqrt{\beta^2_F} \, ,
\end{equation}
where $\bar k_{F,E}$ and $\bar k_{F,B}$ are specific line-of-sight-averaged contractions of $(k_F)^{\alpha\beta\mu\nu}$ (see \PaperI\ for their explicit expressions), and $\nu$ is replaced by an effective frequency $\nu_f$ (see eq.~(5.5) of \PaperI) when combining datasets from different experiments, following the prescription of~\cite{Gubitosi:2012rg}.


\section{Methodology and datasets}
\label{sec:methodology}
We perform a Markov chain Monte Carlo (MCMC) analysis using \texttt{Cobaya}~\cite{Torrado:2020dgo}, jointly sampling the six $\Lambda$CDM parameters\footnote{The six standard $\Lambda$CDM parameters are the baryon and cold dark matter densities $\omega_b \equiv \Omega_b h^2$ and $\omega_c \equiv \Omega_c h^2$, the angular size of the sound horizon at recombination $100\,\theta_{\rm MC}$, the reionization optical depth $\tau$, and the amplitude and tilt of the primordial scalar power spectrum, $\ln(10^{10} A_s)$ and $n_s$.}, the tensor-to-scalar ratio $r$, and the Lorentz-violating parameters $\beta^2_{AF,T}$, $\beta^2_{AF,S}$ and $\beta^2_F$, for both current data and forecasts. We assume a spatially flat Universe with adiabatic initial conditions, one massive and two massless neutrino species with $\sum m_\nu = 0.06\,{\rm eV}$~\cite{deSalas:2020pgw,Esteban:2020cvm,Capozzi:2021fjo}, and effective number of relativistic species $N_{\rm eff} = 3.044$~\cite{Mangano:2001iu,Bennett:2019ewm,Bennett:2020zkv,Akita:2020szl,Froustey:2020mcq,Drewes:2024wbw}. 
We adopt uniform priors $\beta^2_{AF,T},\,\beta^2_{AF,S}\in[0,\,0.01]$ and $\beta^2_F\in[0,\,0.1]$ on the LV parameters, and $r\in[0,1]$ on the tensor-to-scalar ratio, while for the $\Lambda$CDM parameters we follow~\cite{Planck:2018vyg}.
Convergence of the chains is assessed via the Gelman-Rubin criterion~\cite{Gelman:1992zz}, requiring $R - 1 \lesssim 0.01$.
The datasets and analysis methodology for current CMB experiments and the Simons Observatory and \textit{LiteBIRD} forecasts are described below.

\subsection{Current CMB data}
\label{sec:current_data}
We consider different dataset combinations, in order to assess the impact of different CMB observables on the constraints on Lorentz-violating parameters. The labels of the individual datasets are summarized in table~\ref{tab:datasets}.

\paragraph{Temperature and linear polarization.}
As a reference point for comparison, we repeat the analysis of \PaperI, with minor changes:\footnote{With respect to \PaperI, we make three changes: we use the \texttt{lite} version of the \texttt{Plik} likelihood instead of the multi-frequency one, as the former is also used in the combined \textit{Planck}+ACT+SPT analysis; we replace the large-scale $E$-mode polarization likelihood with \texttt{SRoll2}; and we adopt \textit{Planck} PR4 lensing reconstruction instead of the 2018 legacy one.}
\begin{itemize}
    \item \textbf{P18+BK18}: for \textit{Planck}, we use large-scale temperature and $E$-mode polarization data from the \texttt{Commander}~\cite{Planck:2018vyg,Planck:2019nip} and \texttt{SRoll2}~\cite{Pagano:2019tci,Delouis:2019bub} likelihoods, respectively, complemented at high multipoles by the \texttt{Plik\_lite} 
    likelihood~\cite{Planck:2018vyg}. We also include lensing reconstruction measurements from \textit{Planck} PR4~\cite{Carron:2022eyg}. For $B$ modes, we use data from BICEP/\textit{Keck} 2018 (BK18)~\cite{BICEP:2021xfz}.
\end{itemize}

We then consider two additional combinations, obtained by adding new temperature and linear-polarization data from the ACT and SPT experiments, with the corresponding multipole cuts summarized below:
\begin{itemize}
    \item \textbf{SPA+BK18}: we combine \textit{Planck}, ACT DR6 and SPT-3G D1 temperature, polarization and lensing data, which we collectively refer to as SPA, using the \texttt{ACT-lite}~\cite{AtacamaCosmologyTelescope:2025blo} and \texttt{SPT-lite}~\cite{SPT-3G:2025bzu,Balkenhol:2024sbv} likelihoods. 
    Following the strategy adopted in~\cite{AtacamaCosmologyTelescope:2025blo,SPT-3G:2025bzu}, we restrict {\it Planck} to multipoles $\ell \le 1000$ in temperature and $\ell \le 600$ in polarization to minimize the correlations with ACT, and we neglect correlations between SPT and the other datasets.
    We further include the joint \textit{Planck}+ACT DR6 lensing likelihood~\cite{ACT:2023kun,ACT:2023dou,ACT:2023ubw,Carron:2022eyg} and the MUSE lensing likelihood for SPT-3G D1~\cite{SPT-3G:2024atg}. This combination allows us to assess the impact of improved small-scale measurements on the constraints;
    \item \textbf{SPA+BK18+SPT$\boldsymbol{BB}$+SPTpol$\boldsymbol{BB}$ (baseline)}: we further add $B$-mode measurements from SPT-3G D1~\cite{SPT-3G:2025vtb}, restricted to $350 < \ell < 450$, and from 
    SPTpol~\cite{SPT:2019nip}, restricted to $\ell > 450$, to avoid overlap between the datasets (see left panel of figure~\ref{fig:CMB_spectra}). For both we use the \texttt{candl} implementation~\cite{Balkenhol:2024sbv} of the \texttt{lite} likelihoods, introduced in~\cite{Balkenhol:2025xas}. This combination allows us to assess the impact of additional $B$-mode data.
\end{itemize}

\paragraph{$\boldsymbol{EB}$ spectrum.}
We include the parity-violating $EB$ spectrum using the \texttt{Lolli\-pop} likelihood~\cite{Tristram:2020wbi, Tristram:2021tvh}, based on the \textit{Planck}~PR4 data release, published in 2020 (hence the label P20).
Since \texttt{Lollipop} provides the low-$\ell$ $EE$, $BB$ and $EB$ spectra jointly, whenever we use the $EB$ spectrum we consistently include the corresponding \texttt{Lollipop} $EE$ and $BB$ spectra as well. In these combinations, \texttt{Lollipop} therefore replaces the \texttt{SRoll2} low-$\ell$ $E$-mode likelihood. To maximize the information coming from these low-$\ell$ polarization spectra while avoiding any double counting, we use \texttt{Lollipop} up to $\ell = 150$ and correspondingly raise the minimum multipole of the high-$\ell$ \texttt{Plik\_lite} $EE$ spectrum to $\ell_{\min} = 151$. We further neglect the correlations between \textit{Planck} and BK18 $B$-mode data, as done in~\cite{Galloni:2024lre}.
Through eq.~\eqref{eq:EB}, the $EB$ spectrum provides direct sensitivity to isotropic birefringence, i.e.\ to $\beta^2_{AF,T}$, complementing the indirect constraint coming from the  $E$-to-$B$ leakage in the $BB$ spectrum.\footnote{A $TB$ signal is also predicted (see eq.~\eqref{eq:TB}), but we do not include $TB$ data. Besides carrying less statistical weight, the temperature field is more strongly affected by foreground contamination than the $E$-mode field, making $EB$ the cleaner and more constraining parity-violating probe.} We include the $EB$ spectrum in two dataset combinations:
\begin{itemize}
    \item \textbf{P18+P20$\boldsymbol{EB}$+BK18}: the P18+BK18 combination together with the \texttt{Lollipop} spectra described above;
    \item \textbf{SPA+P20$\boldsymbol{EB}$+BK18+SPT$\boldsymbol{BB}$+SPTpol$\boldsymbol{BB}$}: the baseline combination together with the \texttt{Lollipop} spectra described above. 
\end{itemize}
These combinations allow us to assess the constraining power of the direct parity-violating observable.

\paragraph{Circular polarization.}
In \PaperI\ we showed that the $V$-mode measurements from two years of CLASS observations at 40~GHz~\cite{Padilla:2019dhz} and from SPIDER~\cite{SPIDER:2017kwu} do not improve the constraint on the CPT-even parameter $\beta^2_F$. Here we use the latest CLASS results~\cite{Essinger-Hileman:2026tgi}, which extend the measurements to 90, 150 and 220~GHz. We use the minimum-variance combination of the three frequency bands, in bins of $\Delta\ell=20$ over $30\le\ell\le690$.\footnote{The bandpowers are not publicly released. We extract them, together with their uncertainties, from the vector graphics of figure~6 of~\cite{Essinger-Hileman:2026tgi}. Applying the same procedure to the single-frequency spectra of their figure~5, we reproduce the $\chi^2$ values reported there. Since the bandpower covariance is not available either, we neglect correlations between bins, which are expected to be small given the bin width and the large sky fraction observed by CLASS.} The limits from CLASS at 40~GHz, based on observations up to 2022~\cite{Eimer:2023esh}, and from SPIDER are about, respectively, one and three-to-four orders of magnitude weaker (see right panel of figure~\ref{fig:CMB_spectra}), and we do not include them. Following \PaperI, we build a simple Gaussian likelihood
\begin{equation}
    -2 \ln\mathcal{L} = \sum_b \frac{(\mathcal{D}^{VV}_{b,\mathrm{th}} - \mathcal{D}^{VV}_{b,\mathrm{data}})^2}{\sigma^2_b} \, ,
\end{equation}
where $\mathcal{D}^{VV}_{b,\mathrm{th}}$ and $\mathcal{D}^{VV}_{b,\mathrm{data}}$ are the binned theoretical and measured $VV$ bandpowers and $\sigma_b$ the associated uncertainty. We include the $V$-mode spectrum in one dataset combination:
\begin{itemize}
    \item \textbf{SPA+BK18+SPT$\boldsymbol{BB}$+SPTpol$\boldsymbol{BB}$+CLASS}: the baseline combination together with CLASS data. This combination allows us to assess the impact of circular polarization on the constraints.
\end{itemize}

\begin{table}[!t]
\centering
\renewcommand{\arraystretch}{1.3}
\resizebox{0.8\textwidth}{!}{%
\begin{tabular}{l l}
\hline\hline
\textbf{Label} & \textbf{Data} \\
\hline
P18 & \textit{Planck} 2018 $TT$, $TE$, $EE$~\cite{Planck:2018vyg,Planck:2019nip,Pagano:2019tci,Delouis:2019bub}; \textit{Planck} PR4 lensing~\cite{Carron:2022eyg} \\
SPA & \textit{Planck} 2018, ACT DR6~\cite{AtacamaCosmologyTelescope:2025blo} and SPT-3G D1~\cite{SPT-3G:2025bzu,Balkenhol:2024sbv} $TT$, $TE$, $EE$; \\
 & \textit{Planck}+ACT DR6~\cite{Carron:2022eyg,ACT:2023kun,ACT:2023dou,ACT:2023ubw} and SPT-3G D1~\cite{SPT-3G:2024atg} lensing \\
BK18 & BICEP/\textit{Keck} 2018 $BB$~\cite{BICEP:2021xfz} \\
SPT$BB$ & SPT-3G D1 $BB$~\cite{SPT-3G:2025vtb,Balkenhol:2024sbv,Balkenhol:2025xas} \\
SPTpol$BB$ & SPTpol $BB$~\cite{SPT:2019nip,Balkenhol:2024sbv,Balkenhol:2025xas} \\
P20$EB$ & \textit{Planck} PR4 $EE$, $BB$, $EB$, replacing \texttt{SRoll2}~\cite{Tristram:2020wbi,Tristram:2021tvh} \\
CLASS & CLASS 90/150/220\,GHz combined $VV$~\cite{Essinger-Hileman:2026tgi} \\
\hline\hline
\end{tabular}
}
\caption{Labels of the individual datasets used in this work. Dataset combinations with the corresponding multipole cuts are described in section~\ref{sec:current_data}.}
\label{tab:datasets}
\end{table}

\subsection{Forecasts for future CMB experiments}
\label{sec:forecasts}
We forecast the sensitivity of the Simons Observatory (SO) and the \textit{LiteBIRD} satellite to LV parameters. For both experiments, we employ the Wishart likelihood~\cite{Perotto:2006rj, Gerbino:2019okg} as implemented in \texttt{Cobaya} in~\cite{Rashkovetskyi:2021rwg},\footnote{\url{https://github.com/misharash/cobaya_mock_cmb}} which we extend to include $V$ modes.
For Gaussian CMB fields it is exact on the full sky; partial sky coverage is approximated by  scaling the effective number of degrees of freedom per multipole by $f_{\rm sky}$. The likelihood reads
\begin{equation}
    \label{eq:wishart_lkl}
    -2\ln\mathcal{L} = \sum_\ell (2\ell+1)\, f_{\rm sky} 
    \left[ \mathrm{tr}\!\left(\bar{C}_\ell^{-1}\hat{C}_\ell\right) + \ln\frac{|\bar{C}_\ell|}{|\hat{C}_\ell|} - n \right] \, ,
\end{equation}
where $\bar{C}_\ell$ and $\hat{C}_\ell$ are, respectively, the theoretical (signal plus noise) and observed covariance matrices of the $n$ fields, and  $|\,\cdot\,|$ denotes their determinant. Our analysis includes $n=4$ fields, $\{T,E,B,V\}$. We do not include the $EB$ and $TB$ cross-spectra, so that the covariance matrix is block-diagonal, with the $TT$, $TE$ and $EE$ spectra forming one block and the $BB$ and $VV$ spectra entering independently, each over its own multipole range. Our forecasts on $\beta^2_{AF,T}$ are therefore directly comparable with the current-data combinations that do not include the $EB$ spectrum. 

When noise curves are not publicly available, they are modelled as white noise with a Gaussian beam (see e.g.~\cite{Perotto:2006rj,Wu:2014hta,Gerbino:2019okg}):
\begin{equation}
    \label{eq:noise}
    N_\ell^{XX} = \sigma_X^2 \exp\left[ \frac{\ell(\ell+1) \theta_{\rm FWHM}^2}{8 \ln 2} \right] \, , 
\end{equation}
with $X \in \{T, E, B, V \}$, where $\sigma_X$ denotes the instrumental sensitivity and $\theta_{\rm FWHM}$ the full-width half-maximum beam size. We assume that temperature and polarization instrumental noises are uncorrelated, so that $N_\ell^{TE} = 0$.
We adopt the following noise configurations:
\begin{itemize}
    \item \textbf{\textit{Planck}$+$SO}:  we model the noise properties of SO following the specifications provided by the collaboration~\cite{SimonsObservatory:2018koc,SimonsObservatory:2025avm}. We include both the Small Aperture Telescopes (SATs), for $B$-mode polarization, and the Large Aperture Telescope (LAT), for temperature and $E$-mode polarization. 
    For \textit{Planck}, we construct the noise curves via eq.~\eqref{eq:noise}, assuming a single effective channel with $40\,\mu$K-arcmin in temperature and $80\,\mu$K-arcmin in $E$-mode polarization, and a beam size $\theta_{\rm FWHM}=7$ arcmin, as in~\cite{Raffuzzi:2024wyh}.\footnote{We have verified that this noise model reproduces the bounds on the LV parameters obtained using real \textit{Planck} 2018 data.}
    The two experiments are combined as follows (see~\cite{Caloni:2026dyu,Raffuzzi:2024wyh}).
    At $2 \le \ell \le 50$, over $f_{\rm sky} = 0.7$, we use \textit{Planck} alone.
    At higher multipoles, in the sky region observed by SO LAT ($f_{\rm sky} = 0.4$) we adopt the combined SO LAT+\textit{Planck} foreground-marginalized noise curves provided by the SO collaboration for the baseline noise level,\footnote{\url{https://github.com/simonsobs/so_noise_models}} used for $51 \le \ell \le 3000$. To account for the sky area not covered by SO LAT, we include an additional \textit{Planck}-only contribution with $f_{\rm sky} = 0.3$, used at $51 \le \ell \le 2500$. 
    For $B$-mode polarization, we use the SATs over $30 \le \ell \le 300$, with $f_{\rm sky} = 0.1$. Since no component-separated noise curves are publicly available for the SATs, we compute the noise power spectrum of each frequency channel with the SO noise model, for baseline sensitivity and pessimistic $1/f$ noise (see table~2 of~\cite{SimonsObservatory:2018koc}), and combine the channels with inverse-noise weighting. This corresponds to the ideal case without foreground residuals.
    We do not apply any delensing to the $B$-mode spectrum. A partial removal of the lensing $B$-mode contribution would enhance the constraining power on the tensor-to-scalar ratio $r$ and, potentially, on the LV parameters as well~\cite{SimonsObservatory:2018koc,Raffuzzi:2024wyh}.
    \item \textbf{\textit{LiteBIRD}}: we adopt the instrumental configuration of~\cite{LiteBIRD:2022cnt}.\footnote{The \textit{LiteBIRD} mission concept has recently been revised following the 2024 Mission Definition Review~\cite{LiteBIRD:2026gdf}. The current baseline consists of a single telescope with a larger aperture, providing a better angular resolution at comparable sensitivity, although the detailed configuration is yet to be consolidated. Our forecasts are based on the configuration of~\cite{LiteBIRD:2022cnt}. We expect the constraints on the CPT-odd coefficients, driven by large-scale $B$ modes, to be largely unaffected. Access to temperature and $E$-mode data beyond $\ell\approx600$ could instead tighten the bound on the CPT-even coefficient without circular polarization. The sensitivity to $V$ modes relies on the rotating half-wave plate, which is included in the current baseline configuration but not in the backup one.}
    We consider the frequency channels between 78 and 195 GHz, with sensitivities and beam widths taken from table 13 of~\cite{LiteBIRD:2022cnt}, and assume $\sigma_T = \sigma_{E/B}/\sqrt{2}$. The noise power spectra of the individual channels, obtained via eq.~\eqref{eq:noise}, are then coadded to obtain the noise curves used in the analysis. We do not include foregrounds in the \textit{LiteBIRD} forecasts, which therefore correspond to the ideal case without foreground residuals. We adopt $f_{\rm sky}=0.7$ and include temperature and $E$-mode polarization for $2 \le \ell \le 600$, as well as $B$-mode polarization for $2 \le \ell \le 150$ (see e.g.~\cite{LiteBIRD:2022cnt,LiteBIRD:2023zmo}).
    The $B$-mode noise curve is shown in the left panel of figure~\ref{fig:CMB_spectra}, together with that of the SO SATs.

    In addition to temperature and linear polarization, we explore the sensitivity of \textit{LiteBIRD} to $V$-mode polarization assuming the use of a rotating half-wave plate (HWP) as a polarization modulator.\footnote{The rotating HWP is expected to effectively mitigate the $1/f$ noise at large angular scales~\cite{LiteBIRD:2022cnt,Micheli:2024hfe}, justifying our choice of neglecting its contribution in this work.} While an ideal HWP induces a phase shift of exactly $\pi$ between the two orthogonal components of the incident electromagnetic wave, yielding no sensitivity to circular polarization, realistic non-idealities parametrized by a departure $\beta_{\rm HWP}$ from this ideal value provide sensitivity to $V$-mode signals~\cite{Giardiello:2021uxq,Raffuzzi:2024wyh}. In this work we neglect other effects of HWP non-ideality which can impact the $V$-mode signal, such as differential optical load and cross-polarization~\cite{Giardiello:2021uxq}. The noise levels for linear and circular polarization are related to the total intensity noise $\sigma_T$, as a function of $\beta_{\rm HWP}$, by (see~\cite{Raffuzzi:2024wyh})
    \begin{equation}
        \frac{\sigma_{E/B}}{\sigma_T} = \frac{\sqrt{2}}{\cos^2(\beta_{\rm HWP}/2)} \, , \qquad \frac{\sigma_V}{\sigma_T} = \frac{\sqrt{2}}{\sin\beta_{\rm HWP}} \, ,
    \end{equation}
    from which the noise power spectra follow via eq.~\eqref{eq:noise}. Note that, while a non-zero $\beta_{\rm HWP}$ in principle also degrades the sensitivity to linear polarization, this effect is small for values of $\beta_{\rm HWP}$ of the order of a few to a few tens of degrees~\cite{LiteBIRD:2026qdz, Komatsu034005, Pisano:2020qxq} (see also the discussion in section~\ref{sec:litebird}).
    In our analysis we consider four representative values, $\beta_{\rm HWP} = 1^\circ$, $5^\circ$, $10^\circ$, and $20^\circ$, and include $V$-mode polarization for $2 \le \ell \le 600$. 
    The $V$-mode noise curves, for the different values of $\beta_{\rm HWP}$, are shown in the right panel of figure~\ref{fig:CMB_spectra}.
\end{itemize}

As our fiducial cosmology, we adopt the mean values of the $\Lambda$CDM parameters obtained from our analysis with SPA+BK18+SPT$BB$+SPTpol$BB$, setting the tensor-to-scalar ratio and all the LV parameters to zero.

It is worth stressing that our forecasts are simplified and are meant to illustrate a best-case scenario. Apart from the component separation included in the SO LAT noise curves, we neglect foreground residuals and instrumental systematics. They are therefore not intended to represent the realistic sensitivity of SO and \textit{LiteBIRD}.


\section{Results}
\label{sec:results}
In this section we present our results on the LV parameters $\beta^2_{AF,T}$, $\beta^2_{AF,S}$ and $\beta^2_F$. We report 95\% credible level (CL) upper limits unless otherwise stated.
Section~\ref{sec:current_data_results} reports the constraints from current data (table~\ref{tab:current}), while sections~\ref{sec:SO} and~\ref{sec:litebird} present the forecasts for SO and \textit{LiteBIRD}, respectively (table~\ref{tab:forecasts}). Finally, in section~\ref{sec:implicationSME} we recast these results in terms of the SME coefficients.

\begin{table}[!t]
\centering
\renewcommand{\arraystretch}{1.4}
\resizebox{\textwidth}{!}{%
\begin{tabular}{l c c c c}
\hline\hline
\textbf{Dataset} & $r\times10^2$ & $\beta^2_{AF,T}\times10^4$ & $\beta^2_{AF,S}\times10^4$ & $\beta^2_{F}\times10^2$ \\
\hline
P18+BK18                            & $<3.13$ & $<6.77$ & $<6.69$ & $<4.39$ \\
SPA+BK18                            & $<3.12$ & $<6.36$ & $<6.39$ & $<4.70$ \\
SPA+BK18+SPT$BB$+SPTpol$BB$          & $<3.19$ & $<3.59$ & $<3.65$ & $<4.72$ \\
\hdashline
P18+P20$EB$+BK18                     & $<2.80$ & $<7.92$ & $<6.45$ & $<4.80$ \\
SPA+P20$EB$+BK18+SPT$BB$+SPTpol$BB$  & $<2.96$ & $<4.33$ & $<3.46$ & $<5.16$ \\
\hdashline
SPA+BK18+SPT$BB$+SPTpol$BB$+CLASS    & $<3.17$ & $<3.66$ & $<3.82$ & $<4.13$ \\
\hline\hline
\end{tabular}
}
\caption{95\% CL upper limits on $r$ (in units of $10^{-2}$), $\beta^2_{AF,T}$, $\beta^2_{AF,S}$ (in units of $10^{-4}$) and $\beta^2_{F}$ (in units of $10^{-2}$) for the current dataset combinations discussed in section~\ref{sec:current_data_results}.}
\label{tab:current}
\end{table}

\subsection{Constraints from current data}
\label{sec:current_data_results} 
In this section, we discuss the impact of progressively adding the new datasets described in section~\ref{sec:current_data} to \textit{Planck} and BK18 data. We first include the small-scale temperature and $E$-mode measurements from ACT and SPT, then the SPT-3G D1 and SPTpol $B$-mode data, which yield our baseline combination, and the \textit{Planck} $EB$ spectrum. Finally, we discuss the impact of adding the CLASS $V$ modes.

\begin{figure}[!t]
    \centering
    \includegraphics[width=0.49\linewidth]{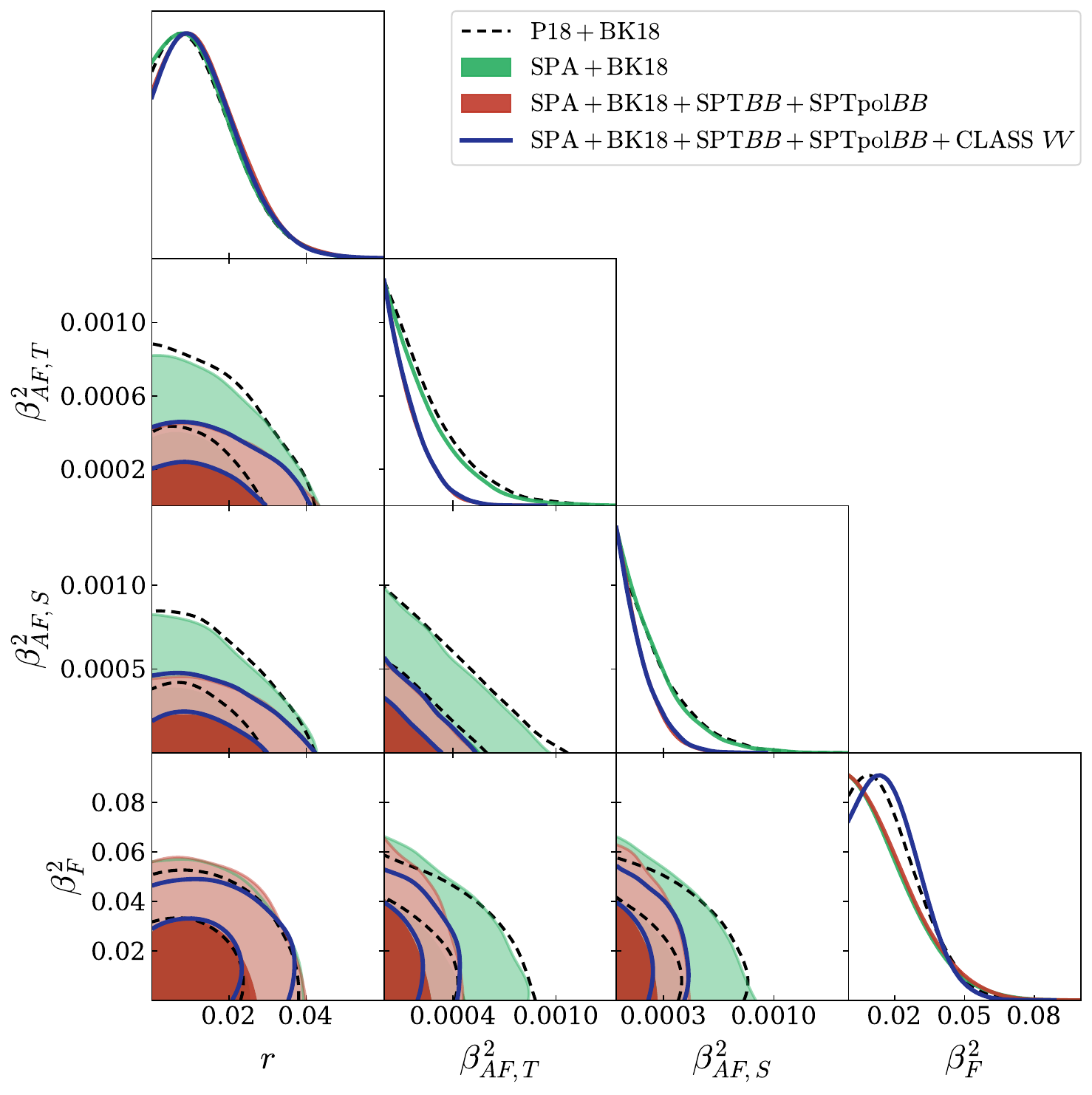}
    \includegraphics[width=0.49\linewidth]{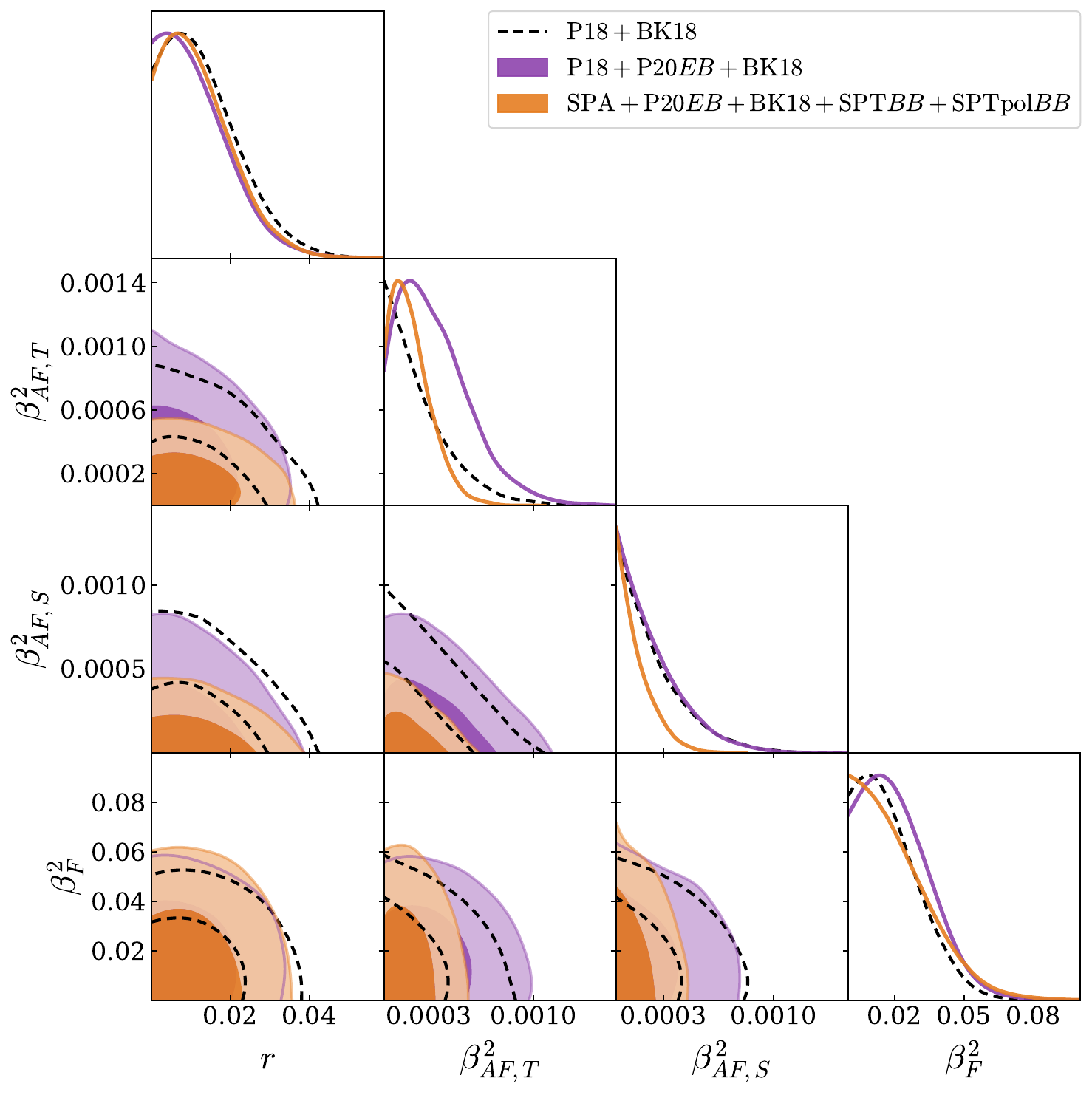}
    \caption{One- and two-dimensional posterior distributions of $r$, $\beta^2_{AF,T}$, $\beta^2_{AF,S}$ and $\beta^2_F$ for different dataset combinations, compared with the P18+BK18 combination (black dashed line) in both panels. \emph{Left}: impact of adding small-scale temperature and polarization data from ACT and SPT (green), and of adding SPT-3G D1 and SPTpol $B$-mode data on top of it (red), and of further adding the CLASS $V$-mode data (dark blue solid line). \emph{Right}: impact of including the \texttt{Lollipop} $EB$ spectrum on top of P18+BK18 (purple) and on top of our baseline dataset combination (orange).}
    \label{fig:posteriors_SPA_Bmodes}
\end{figure}

\paragraph{Small-scale temperature and $\boldsymbol{E}$-mode polarization data.} 
As a first step in our analysis, we test the impact of adding the small-scale temperature and $E$-mode data from ACT and SPT. The 95\% CL bounds on all the LV parameters are only marginally affected, changing by less than $\sim7\%$ with respect to P18+BK18 (table~\ref{tab:current} and left panel of figure~\ref{fig:posteriors_SPA_Bmodes}). For $\beta^2_{AF,T}$ and $\beta^2_{AF,S}$ this is expected, since these parameters are mainly constrained through the leakage of $E$ into $B$ modes probed by BK18, which is included in both combinations, while the additional $E$-mode data are only sensitive to these parameters through the common rescaling factor $\mathcal{Z}$ (eq.~\eqref{eq:Z}). For $\beta^2_F$, which in the absence of $V$-mode data enters the spectra only through $\mathcal{Z}$, the new data could instead be expected to improve the constraint. This does not happen because the scale-independent rescaling by $\mathcal{Z}$ is partially degenerate with the polarization calibration parameters sampled in the ACT and SPT likelihoods, namely the ACT polarization efficiency, $p_{\rm ACT}$, and the SPT $E$-mode calibration, $E^{\rm SPT}_{\rm cal}$, which therefore absorb most of the potential information on $\beta^2_F$. In the baseline combination (SPA+BK18+SPT$BB$+SPTpol$BB$), $\beta^2_F$ has correlation coefficients of $-0.58$ and $-0.38$ with these two parameters, respectively, while the CPT-odd parameters are essentially uncorrelated with them, since their imprint on the $BB$ spectrum cannot be mimicked by a calibration of the polarization spectra (see figure~\ref{fig:calibration_correlations} in appendix~\ref{app:calibration}). Fixing both calibration parameters to unity reduces the width of the $\beta^2_F$ posterior by about 15\%. We perform this test on the baseline combination only, since the $B$-mode data have a negligible impact on the constraint on $\beta^2_F$ (see appendices~\ref{app:B-modes} and~\ref{app:calibration}).

\paragraph{$\boldsymbol{B}$-mode polarization data.}
Next, we add $B$-mode data from SPT-3G D1 and SPTpol. The left panel of figure~\ref{fig:posteriors_SPA_Bmodes} shows that these datasets improve the bounds on $\beta^2_{AF,T}$ and $\beta^2_{AF,S}$ by nearly a factor of 2 with respect to SPA+BK18, while leaving the constraint on $\beta^2_F$ unchanged.  As we show in appendix~\ref{app:B-modes}, this improvement is driven by SPT-3G D1, while SPTpol adds no constraining power and in fact slightly loosens the bounds. Moreover, removing BK18 from the baseline combination weakens the bounds on $\beta^2_{AF,T}$ and $\beta^2_{AF,S}$ by only $\sim15\%$, indicating that current constraints on these parameters rely primarily on the intermediate-scale $B$-mode measurements of SPT-3G D1, which probe the $E$-to-$B$ leakage feature at $\ell\gtrsim300$ (see left panel of figure~\ref{fig:CMB_spectra}), rather than on the $B$-mode measurements of BK18 at larger angular scales.

\paragraph{$\boldsymbol{E}$- and $\boldsymbol{B}$-mode polarization cross-correlation data.}
We then assess the constraining power of the EB spectrum, which we take 
from the \textit{Planck}~PR4 \texttt{Lollipop} likelihood. Unlike the observables considered so far, which probe isotropic birefringence only indirectly through the leakage of $E$ into $B$ modes, the $EB$ spectrum is a direct tracer of this effect (see eq.~\eqref{eq:EB}). 
As a validation of our implementation, we verify that the P18+P20$EB$+BK18 combination with the LV parameters fixed to zero yields $r<0.032$ at 95\% CL, in agreement with~\cite{Tristram:2021tvh}. 

The right panel of figure~\ref{fig:posteriors_SPA_Bmodes} shows that adding the $EB$ spectrum pulls the posterior of $\beta^2_{AF,T}$ about $1\sigma$ away from zero. At 68\% CL, we find $\beta^2_{AF,T}=3.4^{+1.0}_{-3.2}\times10^{-4}$ for P18+P20$EB$\allowbreak+BK18 and $\beta^2_{AF,T}=1.9^{+0.5}_{-1.8}\times10^{-4}$ for SPA+P20$EB$+BK18+SPT$BB$+SPTpol$BB$, while we report the 95\% CL upper limits in table~\ref{tab:current}. 
Moreover, the bound on $r$ tightens with respect to the case with the LV parameters fixed to zero, since $\beta^2_{AF,T}$ and $r$ are partially anticorrelated, both adding power to the $BB$ spectrum.
On the other hand, the bound on $\beta^2_F$ is slightly weakened. This is not due to the $EB$ spectrum itself, but mostly to the replacement of the \texttt{SRoll2} low-$\ell$ $E$-mode likelihood with \texttt{Lollipop}, which changes the constraint on the amplitude of the $EE$ spectrum, and hence on $\mathcal{Z}$. Repeating the P18+P20$EB$+BK18 analysis with only the \texttt{Lollipop} $EE$ spectrum, i.e.\ without the $EB$ and $BB$ spectra, we find a bound on $\beta^2_F$ that already accounts for most of this difference.

Our results for $\beta^2_{AF,T}$ can be compared with existing measurements of the isotropic birefringence angle.
A rotation of the CMB linear polarization plane by an isotropic angle $\alpha$ sources an $EB$ spectrum
\begin{equation}
    C_\ell^{EB}=\frac{1}{2}\sin(4\alpha)\left(\widetilde C_\ell^{EE}-\widetilde C_\ell^{BB}\right),
    \label{eq:EB_rotation}
\end{equation}
where $\widetilde C_\ell^{EE}$ and $\widetilde C_\ell^{BB}$ are the spectra in the absence of rotation~\cite{Lue:1998mq}. This holds regardless of the physical origin of the rotation. Our eq.~\eqref{eq:EB} takes this form with $\sqrt{\beta^2_{AF,T}}=\frac{1}{2}\sin(4\alpha)$. Since $\beta^2_{AF,T}$ is non-negative, our model only describes rotations with $\alpha\geq0$.
A natural benchmark is the value $\alpha=0.36^\circ\pm0.11^\circ$ measured in~\cite{Diego-Palazuelos:2022dsq} from \textit{Planck} PR4 data, the same release on which our $EB$ likelihood is based, but with a substantially different methodology.
We sample the $\Lambda$CDM parameters jointly with all the LV phenomenological parameters, using the full information contained in the CMB spectra.
However, we do not model any residual miscalibration of the detector polarization angles, and thus attribute the entire $EB$ signal to a physical rotation. In~\cite{Diego-Palazuelos:2022dsq}, instead, the cosmological parameters are kept fixed to the \textit{Planck} 2018 $\Lambda$CDM best fit, and $\alpha$ is estimated from the $EB$ cross-spectra of the \textit{Planck} HFI frequency maps jointly with the miscalibration angle of each detector set. The degeneracy between them is broken by exploiting the polarized Galactic foreground emission, which is rotated by a miscalibration but not by cosmic birefringence~\cite{Minami:2019ruj,Minami:2020xfg}.
Galactic dust, however, may carry an intrinsic $EB$ correlation, which would be absorbed into the miscalibration angles and bias $\alpha$ low. The value quoted above includes a correction for this effect, modelled from the $TE$ and $TB$ spectra of dust measured at 353~GHz~\cite{Clark:2021kze}.

\begin{figure}
    \centering
    \includegraphics[width=0.4\linewidth]{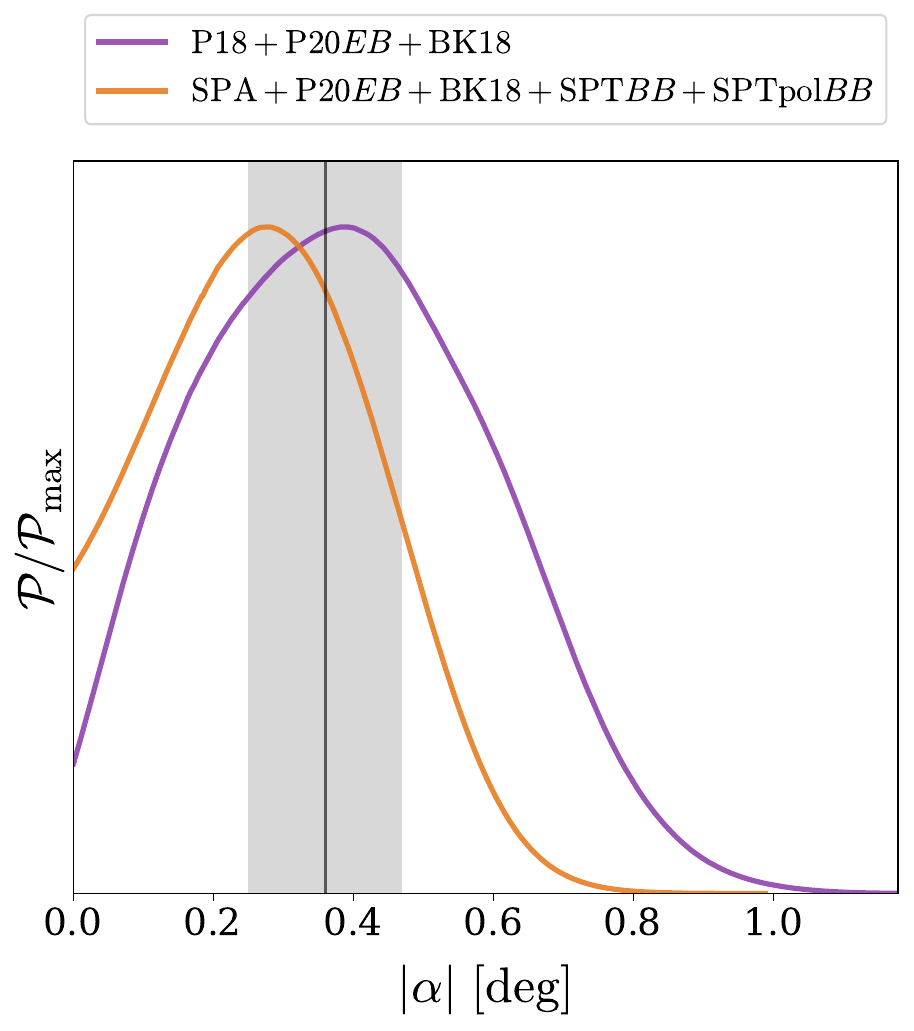}
    \caption{One-dimensional posterior distributions of the isotropic birefringence angle $\alpha$, derived from $\beta^2_{AF,T}$, for the P18+P20$EB$+BK18 (purple) and SPA+P20$EB$+BK18+SPT$BB$+SPTpol$BB$ (orange) dataset combinations, reweighted to a uniform prior on $\alpha$. The black line and the grey band show the central value and the $\pm1\sigma$ region of $\alpha=0.36^\circ\pm0.11^\circ$, measured from \textit{Planck} PR4 data in~\cite{Diego-Palazuelos:2022dsq}, which includes a correction for the intrinsic $EB$ correlation of Galactic dust.}
    \label{fig:alpha_EB_1D}
\end{figure}

Figure~\ref{fig:alpha_EB_1D} shows the posterior distributions of $\alpha$ from the P18+P20$EB$+BK18 and SPA+\allowbreak P20$EB$+BK18+SPT$BB$+SPTpol$BB$ combinations, compared with the measurement of~\cite{Diego-Palazuelos:2022dsq}. Since $\sqrt{\beta^2_{AF,T}}=\frac{1}{2}\sin(4\alpha)$, a uniform prior on $\beta^2_{AF,T}$ corresponds to a prior on $\alpha$ proportional to $\sin(8\alpha)$, which suppresses small rotation angles. For a consistent comparison, we therefore reweight each sample of our chains by $1/\sin(8\alpha)$, which is equivalent to adopting a uniform prior on $\alpha$. The two posteriors peak at about $0.39^\circ$ and $0.27^\circ$, respectively, within the $1\sigma$ region of this measurement. This comparison therefore shows that the mild preference for a non-zero isotropic rotation found in \textit{Planck} PR4 data persists when the cosmological parameters are varied jointly with all the LV parameters. This result should nonetheless be interpreted with caution, since our constraint is sensitive to the total $EB$ signal, including any residual miscalibration of the polarization angles and any residual foreground $EB$ correlation. Disentangling a cosmological rotation from these contributions will require a joint modelling of Lorentz violation and polarization angle miscalibration.

\paragraph{Circular polarization data.}
Finally, we include the CLASS $V$-mode measurements. As discussed in section~\ref{sec:VV_degeneracy}, these data are only sensitive to the sum $\beta^2_F$, which in the previous combinations is constrained only through the rescaling factor $\mathcal{Z}$ (eq.~\eqref{eq:Z}). Adding them to the baseline combination leaves the bounds on $\beta^2_{AF,T}$ and $\beta^2_{AF,S}$ essentially unchanged, while tightening the bound on $\beta^2_F$ by $\sim13\%$ (table~\ref{tab:current} and left panel of figure~\ref{fig:posteriors_SPA_Bmodes}). This improvement arises because, at intermediate angular scales, the CLASS upper limits now lie close to the largest $VV$ spectrum compatible with temperature and linear-polarization data (right panel of figure~\ref{fig:CMB_spectra}), so that $V$-mode data begin to provide constraining power on $\beta^2_F$. Circular polarization also probes this parameter in a cleaner way. While the rescaling by $\mathcal{Z}$ is partially degenerate with the polarization calibration parameters of ACT and SPT, the $VV$ spectrum sourced by the CPT-even operator cannot be mimicked by a miscalibration of the linear polarization. Including CLASS data therefore  reduces the anticorrelation of $\beta^2_F$ with these calibration parameters (see appendix~\ref{app:calibration}).

\subsection[Simons Observatory: impact of small-scale temperature and $E$-mode data]{Simons Observatory: impact of small-scale temperature and $\boldsymbol{E}$-mode data}
\label{sec:SO}
We first discuss the forecast sensitivity of \textit{Planck}$+$SO to the LV parameters, following the methodology described in section~\ref{sec:forecasts}. As shown in table~\ref{tab:forecasts} and figure~\ref{fig:forecasts}, the largest improvement with respect to current data is on $\beta^2_F$, whose bound tightens by a factor of $\sim4$. This is expected, since in the absence of $V$-mode sensitivity the constraint on $\beta^2_F$ comes entirely from the rescaling of the $TE$ and $EE$ spectra through $\mathcal{Z}$ (eq.~\eqref{eq:Z}), and therefore benefits directly from the high-precision small-scale temperature and $E$-mode measurements of SO LAT. Nevertheless, we note that our forecasts do not include polarization calibration uncertainties, which, as found for current data, are partially degenerate with $\beta^2_F$ (see appendix~\ref{app:calibration}), so that the forecast improvement on this parameter may be somewhat optimistic. Comparing the widths of the posterior distributions, the improvement with respect to current data analysed with the ACT and SPT calibration parameters fixed is only slightly smaller, roughly a factor of $3.5$.

The bounds on $\beta^2_{AF,T}$ and $\beta^2_{AF,S}$, instead, improve by a factor of $\sim1.5$ with respect to current data. This improvement comes from the higher sensitivity of the SO SATs with respect to BK18 over the same range of angular scales, $30\le\ell\le300$. These data, however, do not cover the $E$-to-$B$ leakage feature at $\ell\gtrsim300$, where most of the information on these two parameters is contained (see section~\ref{sec:current_data_results} and appendix~\ref{app:B-modes}). $B$-mode measurements at smaller angular scales than those probed by the SATs, e.g. with SO LAT~\cite{SimonsObservatory:2018koc}, would therefore further tighten the bounds on $\beta^2_{AF,T}$ and $\beta^2_{AF,S}$.
Although $r$ is always varied in the analysis, we do not show it in table~\ref{tab:forecasts} and figure~\ref{fig:forecasts}, since its constraint is driven by the standard $B$-mode sensitivity of the experiment. 
For reference, when the LV parameters are varied, we find $r<3.5\times10^{-3}$ at 95\% CL, about an order of magnitude tighter than with current data (see table~\ref{tab:current}).\footnote{Fixing the LV parameters to zero, we obtain $r<4\times10^{-3}$ at 95\% CL, in good agreement with the results of~\cite{SimonsObservatory:2018koc} for $\Lambda$CDM$+r$, which, however, also account for the residuals of foreground cleaning.}

\subsection{\textit{LiteBIRD}: the role of future $V$-mode measurements}
\label{sec:litebird}
\begin{figure}[!t]
    \centering
    \includegraphics[width=0.7\linewidth]{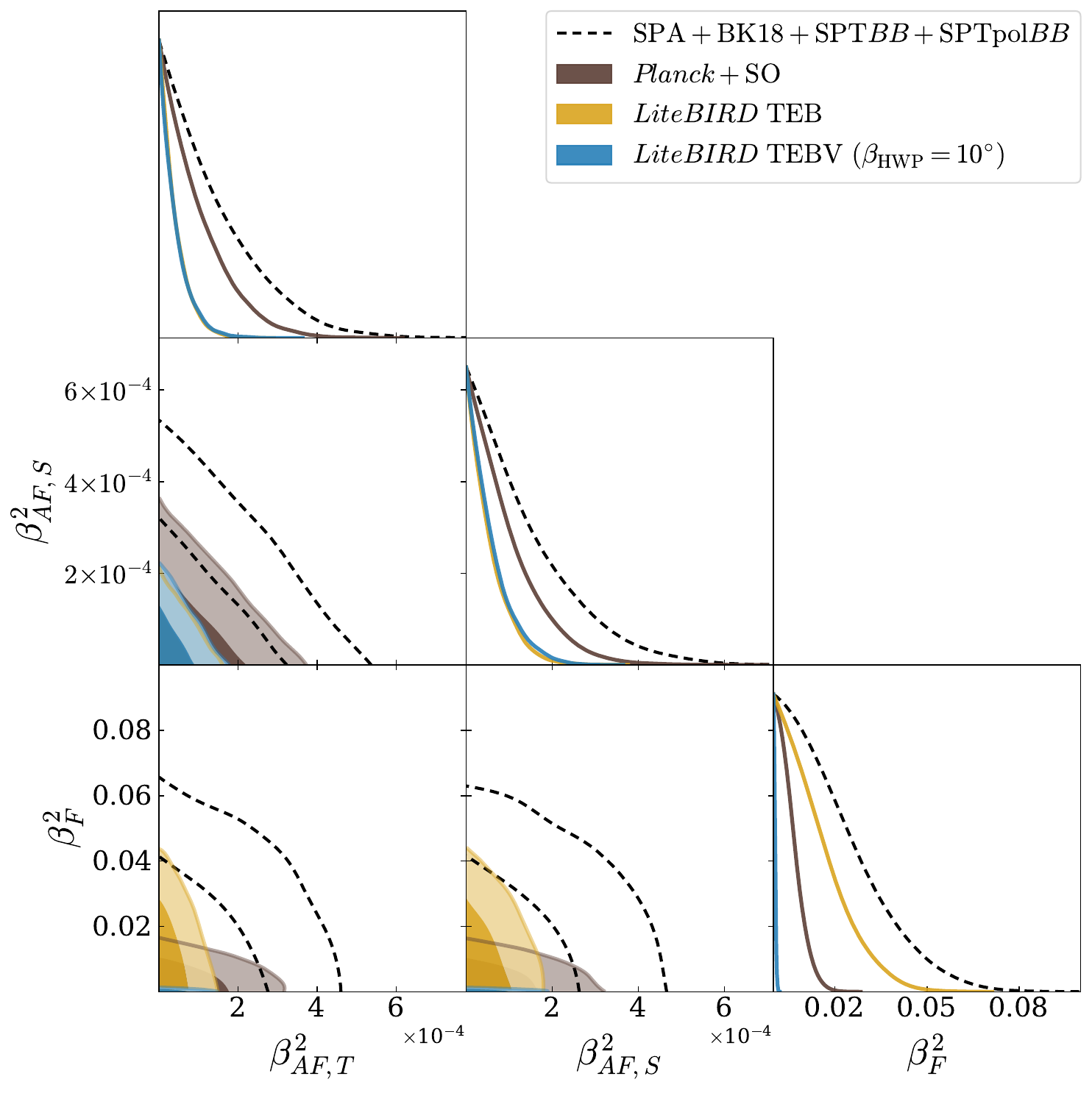}
    \caption{One- and two-dimensional posterior distributions of $\beta^2_{AF,T}$, $\beta^2_{AF,S}$ and $\beta^2_F$ forecast for \textit{Planck}$+$SO (brown), \textit{LiteBIRD} $TEB$ only (yellow) and \textit{LiteBIRD} $TEBV$ with $\beta_{\rm HWP}=10^\circ$ (blue), compared with the constraints from current data, SPA+BK18+SPT$BB$+SPTpol$BB$ (black dashed). The forecasts assume a fiducial cosmology with $r=0$ and all LV parameters set to zero (see section~\ref{sec:forecasts}).}
    \label{fig:forecasts}
\end{figure}

\begin{table}[!t]
\centering
\renewcommand{\arraystretch}{1.4}
\resizebox{0.8\textwidth}{!}{%
\begin{tabular}{l c c c}
\hline\hline
\textbf{Dataset} & $\beta^2_{AF,T}\times10^4$ & $\beta^2_{AF,S}\times10^4$ & $\beta^2_{F}\times10^2$ \\
\hline
\textit{Planck}$+$SO & $<2.48$ & $<2.41$ & $<1.19$ \\
\hdashline
\textit{LiteBIRD} $TEB$ only & $<1.05$ & $<1.38$ & $<3.20$ \\
\hdashline
\textit{LiteBIRD} $TEBV$ ($\beta_{\rm HWP}=1^\circ$)  & $<1.08$ & $<1.42$ & $<2.98$ \\
\textit{LiteBIRD} $TEBV$ ($\beta_{\rm HWP}=5^\circ$)  & $<1.13$ & $<1.50$ & $<0.380$ \\
\textit{LiteBIRD} $TEBV$ ($\beta_{\rm HWP}=10^\circ$) & $<1.12$ & $<1.50$ & $<0.0954$ \\
\textit{LiteBIRD} $TEBV$ ($\beta_{\rm HWP}=20^\circ$) & $<1.16$ & $<1.48$ & $<0.0246$ \\
\hline
SPA+BK18+SPT$BB$+SPTpol$BB$ & $<3.59$ & $<3.65$ & $<4.72$ \\
\hline\hline
\end{tabular}
}
\caption{95\% CL upper limits on $\beta^2_{AF,T}$, $\beta^2_{AF,S}$ (in units of $10^{-4}$) and $\beta^2_{F}$ (in units of $10^{-2}$) for the \textit{Planck}$+$SO and \textit{LiteBIRD} forecasts. For the latter, we report results without ($TEB$) and with ($TEBV$) sensitivity to circular polarization at different values of the HWP phase shift $\beta_{\rm HWP}$. For comparison we also include the bounds obtained with current observations, SPA+BK18+SPT$BB$+SPTpol$BB$.}
\label{tab:forecasts}
\end{table}
We now turn to the forecast sensitivity of the \textit{LiteBIRD} satellite, first considering a $TEB$-only configuration, i.e.\ without any sensitivity to circular polarization. As shown in table~\ref{tab:forecasts} and figure~\ref{fig:forecasts}, \textit{LiteBIRD} improves the bounds on $\beta^2_{AF,T}$ and $\beta^2_{AF,S}$ by a factor of $\sim3.4$ and $\sim2.6$ with respect to current data, and by a factor of $\sim2.4$ and $\sim1.7$ with respect to \textit{Planck}$+$SO. This improvement is driven by the high-sensitivity B-mode measurements of LiteBIRD at the largest angular scales, which are not accessible to SO and where the Planck noise is too high to be informative. There, the $E$-to-$B$ leakage produces an excess in the $BB$ spectrum even for $r=0$ (see left panel of figure~\ref{fig:CMB_spectra}).
Moreover, at these scales the leakage sourced by $\beta^2_{AF,S}$ is weaker than that sourced by $\beta^2_{AF,T}$. In eq.~\eqref{eq:K22}, $\beta^2_{AF,T}$ multiplies $\widetilde C^{EE}_\ell$ with a unit coefficient, while $\beta^2_{AF,S}$ multiplies $\widetilde C^{EE}_{\ell-1}$ and $\widetilde C^{EE}_{\ell+1}$ with weights $(\ell^2-4)/[\ell(2\ell+1)]$ and $(\ell-1)(\ell+3)/[(\ell+1)(2\ell+1)]$, whose sum approaches unity only at high multipoles and drops to $1/3$ at $\ell=2$.
This explains the larger improvement on $\beta^2_{AF,T}$. The bound on $\beta^2_F$, instead, improves only by a factor of $\sim1.5$ with respect to current data, and is nearly 3 times weaker than the \textit{Planck}$+$SO one. As for current data, without $V$-mode sensitivity the constraint on $\beta^2_F$ comes entirely from the rescaling of the $TE$ and $EE$ spectra through $\mathcal{Z}$, to which \textit{LiteBIRD}, limited to $\ell\le600$, is less sensitive than SO. In the $TEB$-only case, the two experiments are therefore complementary, with SO dominating the constraint on the CPT-even parameter and \textit{LiteBIRD} those on the CPT-odd ones.

Including $V$-mode sensitivity through a non-ideal HWP changes this picture for $\beta^2_F$. As $\beta_{\rm HWP}$ increases from $1^\circ$ to $20^\circ$, the bound on $\beta^2_F$ tightens by two orders of magnitude, eventually becoming comparable to those on $\beta^2_{AF,T}$ and $\beta^2_{AF,S}$. This trend can be understood from the right panel of figure~\ref{fig:CMB_spectra}: larger values of $\beta_{\rm HWP}$ lower the $V$-mode noise, improving the signal-to-noise ratio of the predicted $VV$ spectrum and hence the sensitivity to $\beta^2_F$. For $\beta_{\rm HWP}=1^\circ$, $V$-mode data tighten the bound by only $\sim7\%$ with respect to the $TEB$-only case, a contribution similar to that of current CLASS data (see section~\ref{sec:current_data_results}). Already for $\beta_{\rm HWP}=5^\circ$, however, the \textit{LiteBIRD} bound on $\beta^2_F$ is about an order of magnitude tighter than the current one including CLASS, and a factor of $\sim3$ tighter than the \textit{Planck}$+$SO one. The bounds on $\beta^2_{AF,T}$ and $\beta^2_{AF,S}$, on the other hand, remain stable within $\sim10\%$ across all values of $\beta_{\rm HWP}$ considered, as do the uncertainties on the standard $\Lambda$CDM$+r$ parameters, with a 95\% CL upper limit $r\lesssim4\times10^{-4}$ in all cases when the LV parameters are varied.\footnote{Fixing the LV parameters to zero, we obtain a comparable bound on $r$. This is tighter than the one reported in~\cite{LiteBIRD:2022cnt}, which accounts for foreground residuals and instrumental systematics, both neglected in our forecasts (see section~\ref{sec:forecasts}). Our bound with the LV parameters varied is instead in agreement with the one obtained in~\cite{Raffuzzi:2024wyh} with a similar approach, based only on white noise and beam, for an analogous extension of $\Lambda$CDM.} A non-ideal HWP therefore improves the constraint on $\beta^2_F$ without degrading the sensitivity to the other parameters, as already noted in~\cite{Raffuzzi:2024wyh}.

\subsection{Implications for the minimal SME coefficients}
\label{sec:implicationSME}
In this section, we recast the constraints on the phenomenological parameters $\beta^2_{AF,T}$, $\beta^2_{AF,S}$ and $\beta^2_F$ presented above in terms of the fundamental coefficients appearing in the SME action, eq.~\eqref{eq:action-LV}, using the mapping described in section~\ref{sec:connections-beta-k}. The results are reported in table~\ref{tab:bounds_sme_coefficients}.
For the baseline combination, SPA+BK18+SPT$BB$+SPTpol$BB$, we find at 95\% CL:
\begin{align}
    |k^{(3)}_{(V)00}| &< 7.75\times10^{-45}\; \text{GeV} \, , \\
    |\mathbf{k_{AF}}| &< 3.82\times10^{-45}\; \text{GeV} \, , \\
    k_{F,E+B} &< 5.99\times10^{-32}\left(\frac{\nu_f}{158.8\; \text{GHz}}\right)^{-1} \, .
\end{align}
Note that the bound on $k_{F,E+B}$ is normalized to the effective frequency of the \textit{Planck} channels, $\nu_f=158.8$ GHz, since the constraint on $\beta^2_F$ is driven by $EE$ and $TE$ \textit{Planck} data.\footnote{As discussed in appendix~\ref{app:calibration}, the \texttt{Plik\_lite} likelihood keeps the \textit{Planck} polarization efficiencies fixed, while the ACT and SPT \texttt{lite} likelihoods each sample their own polarization calibration parameter, which is degenerate with $\beta^2_F$. ACT and SPT polarization data therefore add little constraining power on this parameter.} 
This value is computed using the sensitivities and nominal bandwidths of the 100, 143 and 217 GHz channels~\cite{Planck:2018nkj}. Following \PaperI, the \textit{Planck} channels are weighted by their temperature sensitivities, since the constraint on $\beta^2_F$ comes also from the $TE$ spectrum. Weighting them by their polarization sensitivities would instead give $\nu_f\simeq151.3$~GHz, loosening the bound on $k_{F,E+B}$ by about 5\%.
 
Compared to the constraints of \PaperI, obtained with \textit{Planck}+BK18+ACT DR4 data, the bounds on both CPT-odd coefficients are reduced by $\sim25\%$.
These constraints represent the strongest bounds on the CPT-odd coefficients of the minimal SME to date, considering both CMB and non-CMB probes (see~\cite{Kostelecky:2008ts} for a compilation of existing constraints).
When the \textit{Planck} PR4 $EB$ spectrum is included, the posterior of $|k^{(3)}_{(V)00}|$ peaks away from zero, reflecting the mild preference for a non-zero isotropic rotation discussed in section~\ref{sec:current_data_results}, so that the 95\% CL upper limit shifts to $8.51\times10^{-45}$\,GeV. The bound on $|\mathbf{k_{AF}}|$ is instead essentially unchanged.
Looking at the forecasts, \textit{Planck}$+$SO would further reduce the bounds on the CPT-odd SME coefficients by $\sim20\%$ with respect to current data, while \textit{LiteBIRD} in the $TEB$-only configuration would reduce them by $\sim45\%$ for $|k^{(3)}_{(V)00}|$ and $\sim40\%$ for $|\mathbf{k_{AF}}|$.
Including $V$-mode sensitivity through a non-ideal HWP has no relevant impact on these coefficients, as the bounds change by at most $\sim 5\%$ with respect to the $TEB$-only case for all the values of $\beta_{\rm HWP}$ considered.

For the CPT-even coefficient, our baseline bound on $k_{F,E+B}$ is comparable to that of \PaperI. Including CLASS data tightens it by about 7\%.
Note that the bound we report with CLASS is normalized to the effective frequency of \textit{Planck}, which continues to provide most of the constraining power on $k_{F,E+B}$.
Astrophysical sources emitting more energetic photons provide much stronger bounds on the CPT-even coefficients, from the $10^{-34}$--$10^{-35}$ level with optical polarimetry~\cite{Friedman:2020bxa,Gerasimov:2021chj} down to $\sim10^{-37}$ with gamma-ray burst polarimetry~\cite{Kostelecky:2006ta,Kostelecky:2013rv}, since the effect of this operator grows with the photon energy. These works, however, constrain different combinations of the components of $(k_F)^{\alpha\beta\mu\nu}$ from the one probed by our analysis, $k_{F,E+B}$, so the two kinds of bounds are not directly comparable.

For the forecasts, the effective frequencies are computed from the channels used to construct the noise curves of each experiment. For SO, we use $\nu_f=116.2$ GHz, obtained from the six LAT channels (27, 39, 93, 145, 225 and 280 GHz) with the baseline polarization sensitivities of~\cite{SimonsObservatory:2018koc} and the passbands taken from~\cite{SimonsObservatory:2025wwn}. For \textit{LiteBIRD}, we use $\nu_f=139.6$ GHz, obtained from the seven channels between 78 and 195 GHz (see section~\ref{sec:forecasts}). As reported in table~\ref{tab:bounds_sme_coefficients}, \textit{Planck}$+$SO would reduce the bound on $k_{F,E+B}$ by $\sim 30\%$ with respect to current data, while \textit{LiteBIRD} in the $TEB$-only configuration would reduce it by only $\sim 5\%$, remaining weaker than \textit{Planck}$+$SO. Including $V$-mode sensitivity changes this picture. For $\beta_{\rm HWP}=20^\circ$, the bound improves by a factor of $\sim 12$ with respect to current data and by a factor of $\sim 8$ with respect to \textit{Planck}$+$SO.

\begin{table}[!t]
\centering
\renewcommand{\arraystretch}{1.4}
\resizebox{\textwidth}{!}{%
\begin{tabular}{l c c c}
\hline\hline
\textbf{Dataset} & $|k^{(3)}_{(V)00}| \times 10^{45}$ (GeV) & $|\mathbf{k_{AF}}| \times 10^{45}$ (GeV) & $k_{F,E+B} \times 10^{32}$ \\
\hline
SPA+BK18+SPT$BB$+SPTpol$BB$          & $< 7.75$ & $< 3.82$ & $< 5.99$ \\
SPA+P20$EB$+BK18+SPT$BB$+SPTpol$BB$  & $< 8.51$ & $< 3.72$ & $<6.27$ \\
SPA+BK18+SPT$BB$+SPTpol$BB$+CLASS    & $< 7.82$ & $< 3.91$ & $< 5.61$ \\
\hdashline
\textit{Planck}$+$SO                 & $< 6.44$ & $< 3.10$ & $<4.11$ \\
\hdashline
\textit{LiteBIRD} $TEB$ only                          & $< 4.19$ & $< 2.35$ & $<5.61$ \\
\textit{LiteBIRD} $TEBV$ ($\beta_{\rm HWP}=20^\circ$) & $< 4.40$ & $< 2.43$ & $<0.492$ \\
\hline\hline
\end{tabular}
}
\caption{95\% CL upper limits on the SME coefficients $|k^{(3)}_{(V)00}|$ and $|\mathbf{k_{AF}}|$ (in units of $10^{-45}$\,GeV) and $k_{F,E+B}$ (in units of $10^{-32}$), obtained by recasting the bounds on $\beta^2_{AF,T}$, $\beta^2_{AF,S}$ and $\beta^2_F$ reported in tables~\ref{tab:current} and~\ref{tab:forecasts} through eqs.~\eqref{eq:kv00}--\eqref{eq:kFEB}. 
For $k_{F,E+B}$, the frequency $\nu$ in eq.~\eqref{eq:kFEB} is replaced by an effective frequency $\nu_f$, which is that of \textit{Planck} for current data, including the combination with CLASS, and that of SO and \textit{LiteBIRD} for the respective forecasts (see main text for more details).}
\label{tab:bounds_sme_coefficients}
\end{table}


\section{Conclusions}
\label{sec:conclusion}

In this work, we have updated the CMB constraints on Lorentz-violating electrodynamics within the minimal Standard Model Extension, considering the CPT-odd, dimension-3 operator, which sources cosmic birefringence, and the CPT-even, dimension-4 operator, which converts linear into circular polarization. Combining \textit{Planck}, ACT DR6 and SPT-3G D1 temperature and $E$-mode polarization data with $B$-mode measurements from BICEP/\textit{Keck}, SPT-3G D1 and SPTpol, we obtain $|k^{(3)}_{(V)00}|<7.75\times10^{-45}$\,GeV and $|\mathbf{k_{AF}}|<3.82\times10^{-45}$\,GeV at 95\% CL on the coefficients sourcing isotropic and anisotropic birefringence, respectively. These are the strongest bounds to date on these coefficients from any probe~\cite{Kostelecky:2008ts}. With respect to our previous analysis~\cite{Caloni:2022kwp}, these bounds improve by about 25\%, thanks to the precision of the SPT-3G D1 $B$-mode measurements at $\ell\gtrsim300$, where the leakage of $E$ into $B$ modes induced by these operators is significant. More precise $B$-mode data at these scales would therefore further tighten these constraints. At these scales, the $B$-mode spectrum is dominated by gravitational lensing, so that the $E$-to-$B$ leakage induced by Lorentz violation may also be relevant for lensing analyses, either as a contaminant or, for anisotropic birefringence, through the induced coupling between multipoles.

For the first time in this kind of analysis, we have also included the \textit{Planck} PR4 $EB$ spectrum, which directly traces isotropic birefringence. These data yield a mild preference for a non-zero $k^{(3)}_{(V)00}$, corresponding to a birefringence angle about $1\sigma$ away from zero and consistent, although with a lower significance, with the one measured from the same \textit{Planck} release with a different methodology in~\cite{Diego-Palazuelos:2022dsq}. Since our analysis attributes the entire $EB$ signal to a physical rotation, assessing the origin of this signal will require modelling the miscalibration of the polarization angles jointly with Lorentz violation, together with a better characterization of the polarized Galactic emission used to calibrate them~\cite{Minami:2019ruj,Clark:2021kze}. More generally, several ground-based experiments, including BICEP/\textit{Keck} and SPT-3G, calibrate the global polarization angle by nulling the observed $EB$ and $TB$ spectra~\cite{BICEP:2021xfz,SPT-3G:2025vtb}. This procedure removes any isotropic rotation from the maps, including a cosmological one, so that the corresponding $B$-mode data mainly retain sensitivity to anisotropic birefringence. 
\textit{Planck} and ACT, instead, do not null the $EB$ spectrum~\cite{AtacamaCosmologyTelescope:2025blo,Planck:2016soo}, so that their data retain sensitivity to an isotropic rotation, within the uncertainty on the orientation of their polarization-sensitive detectors~\cite{Planck:2016soo}.
A consistent treatment of the polarization angle calibration across datasets will therefore be essential to robustly constrain the isotropic component.

For the CPT-even coefficient, temperature and linear-polarization data give $k_{F,E+B}<5.99\times10^{-32}$ at 95\% CL. Including the latest CLASS circular polarization measurements at 90, 150 and 220~GHz~\cite{Essinger-Hileman:2026tgi}, we tighten this bound by about 7\%. This is the first time that CMB $V$-mode data contribute to constraining this coefficient. In this respect, we defer to future work the inclusion of the recent data from the POLARBEAR collaboration~\cite{POLARBEAR:2026zav}, which probe angular scales up to $\ell = 2000$. Without circular polarization measurements, the constraint relies on the amplitude rescaling of the linear-polarization spectra, which is degenerate with their calibration factors. An accurate external calibration of the polarization amplitude would therefore make this bound more robust. Furthermore, a multi-frequency analysis, sampling $k_{F,E+B}$ directly rather than relying on an effective frequency, would also fully exploit the frequency dependence of the CPT-even effect.

We have then forecast the sensitivity of Simons Observatory (SO) and \textit{LiteBIRD} to these coefficients. These forecasts neglect foreground residuals in the SO $B$-mode data and in all \textit{LiteBIRD} data, and should therefore be regarded as optimistic. Using linear polarization data only, we find that the two experiments are complementary, with SO tightening the CPT-even bound by about 30\% and \textit{LiteBIRD} the CPT-odd ones by 40 to 45\%. For \textit{LiteBIRD}, we have further considered a configuration including a rotating half-wave plate (HWP), whose non-idealities, parametrized by a departure of its phase shift from the ideal value, provide sensitivity to $V$ modes~\cite{Giardiello:2021uxq,Raffuzzi:2024wyh}. We have explored four representative values of this departure, between $1^\circ$ and $20^\circ$. The CPT-even bound improves rapidly with the phase shift, by more than one order of magnitude for the largest value, without degrading the other constraints. HWP non-idealities, usually treated as a systematic to be mitigated, can thus become a resource for testing fundamental physics. 
We have not considered here the new \textit{LiteBIRD} setup proposed in the revised mission concept~\cite{LiteBIRD:2026gdf}. While we expect the constraints on the CPT-odd coefficients to be largely unaffected, its improved angular resolution could strengthen the CPT-even bound from linear polarization. We defer a quantitative assessment to future work.

A natural extension of this work is to go beyond the CMB power spectra, which are only sensitive to quadratic combinations of the SME coefficients. The direction-dependent components of $k_{AF}$ and $k_F$ induce correlations between different multipoles, $\ell\neq\ell'$, which carry information that power spectra cannot access. For $k_{AF}$, power spectra constrain the amplitude of its spatial part but not its direction, which sets a dipolar modulation of the rotation angle across the sky and could be measured with existing estimators of anisotropic birefringence, see e.g. ~\cite{Zagatti:2024jxm, Gluscevic:2009mm}. For $k_F$, to our knowledge, no analogous estimator for the conversion of linear into circular polarization has been developed yet. Such an estimator would give access to further combinations of the components of $k_F$, separating the two contractions that power spectra only constrain through $k_{F,E+B}$ (eq.~\eqref{eq:kFEB}). We leave such map-level analyses, which will be key to fully exploiting upcoming CMB polarization data as a probe of Lorentz invariance, to future work.

\acknowledgments
We thank Lennart Balkenhol for his support with the SPT data and the use of the \texttt{candl} likelihood framework. We also thank Paolo Campeti, Eric Hivon and Ali Rida Khalife for useful discussions. We acknowledge CINECA for the availability of high performance computing resources and support through the CINECA-INFN agreement for the InDark project. We acknowledge support by Commissione Scientifica Nazionale 4 (CSN4) and the Research Network InDark funded by Istituto Nazionale di Fisica Nucleare (INFN), and by the ASI (Italian Space Agency) Grant 2020-9-HH.0 (participation in LiteBIRD, phase A). We acknowledge the use of \texttt{numpy} \citep{Harris:2020xlr}, \texttt{matplotlib} \citep{Hunter:2007ouj} and \texttt{getdist} \citep{Lewis:2019xzd} software packages. L.C. acknowledges the financial support provided by Fundação para a Ciência e Tecnologia (FCT), I.P., through the Strategic Funding UID/04650/2025 and UID/04564/2025 and national funds with DOI identifiers 10.54499/\allowbreak2023.11681.PEX, 10.54499/2024.00249.CERN funded by measure RE-C06-i06.m02 -- ``Reinforcement of funding for International Partnerships in Science, Technology and Innovation'' of the Recovery and Resilience Plan -- RRP, within the framework of the financing contract signed between the Recover Portugal Mission Structure (EMRP) and the Foundation for Science and Technology I.P. (FCT), as an intermediate beneficiary, as well as the advanced computing projects 2024.00249.CERN.F1 and 2024.07037.CPCA.A1. MG, ML and LP are funded by the European Union (ERC, RELiCS, project number 101116027). Views and opinions expressed are however those of the author(s) only and do not necessarily reflect those of the European Union or the European Research Council Executive Agency. Neither the European Union nor the granting authority can be held responsible for them. This is not an official Simons Observatory or LiteBIRD collaboration paper.

\appendix

\section{Impact of $\boldsymbol{B}$-mode datasets on the constraints}
\label{app:B-modes}

In section~\ref{sec:current_data_results} we showed that adding SPT-3G D1 and SPTpol $B$-mode data to the SPA+BK18 combination improves the bounds on $\beta^2_{AF,T}$ and $\beta^2_{AF,S}$ by nearly a factor of 2. Here we assess the role of $B$-mode data in more detail, first by comparing the baseline combination with the SPA combination alone, and then by removing in turn SPTpol, SPT-3G D1 or BK18 data from the baseline combination.

Without any $B$-mode data, the bounds on $\beta^2_{AF,T}$ and $\beta^2_{AF,S}$ are about 20 times looser than for the baseline combination (table~\ref{tab:bmodes_individual_data}). In this case the CPT-odd parameters are constrained essentially only through the scale-independent rescaling of the $TE$ and $EE$ spectra by $\mathcal{Z}$ (eq.~\eqref{eq:Z}). The comparison with the baseline combination therefore shows that almost all the information on these parameters comes from the $E$-to-$B$ leakage probed by $B$-mode data. Adding BK18 alone already tightens these bounds by a factor of $\sim12$ (table~\ref{tab:current}).

\begin{figure}[!t]
    \centering
    \includegraphics[width=0.7\linewidth]{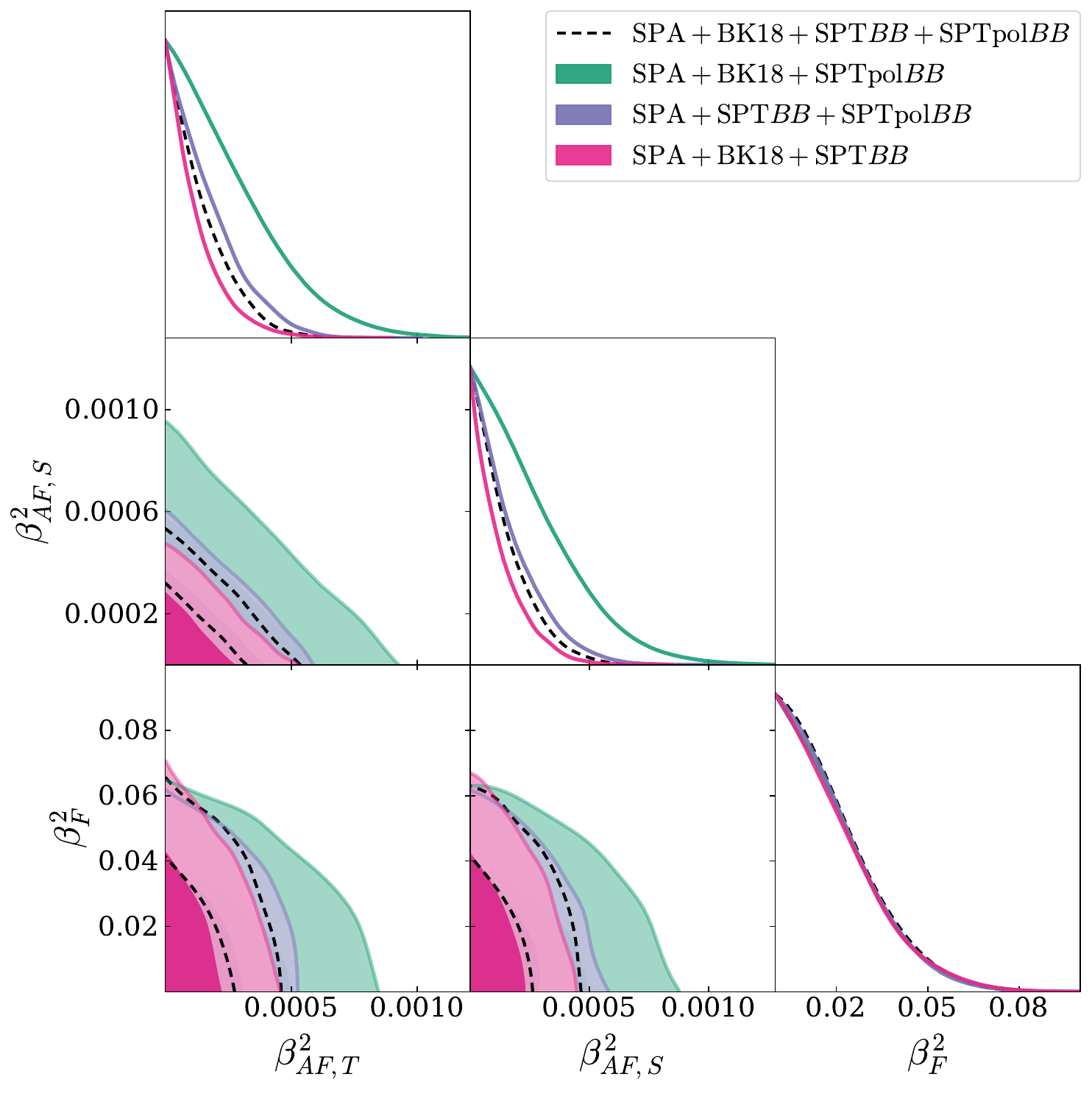}
    \caption{One- and two-dimensional posterior distributions of $\beta^2_{AF,T}$, $\beta^2_{AF,S}$ and $\beta^2_F$ for the baseline combination SPA+BK18+SPT$BB$+SPTpol$BB$ compared with the combinations obtained by removing SPTpol (SPA+BK18+SPT$BB$), SPT-3G D1 (SPA+BK18+SPTpol$BB$) or BK18 (SPA+SPT$BB$+SPTpol$BB$) $B$-mode data.}
    \label{fig:individual_BB}
\end{figure}

The further improvement obtained by adding SPT $B$-mode data to SPA+BK18 is almost entirely driven by SPT-3G D1 (figure~\ref{fig:individual_BB} and table~\ref{tab:bmodes_individual_data}). The SPA+BK18+SPT$BB$ combination yields bounds comparable to, and in fact slightly tighter than, those of the baseline combination, whereas SPA+BK18+SPTpol$BB$ gives bounds roughly a factor of 2 looser on both parameters, close to those of SPA+BK18 (table~\ref{tab:current}). This is likely due to the lowest-multipole SPTpol band power, which lies above the $\Lambda$CDM prediction and therefore prefers a slightly larger $E$-to-$B$ leakage (see left panel of figure~\ref{fig:CMB_spectra}).

We then remove BK18 while keeping both SPT $B$-mode datasets. In this case we also include the SPT-3G D1 band powers that overlap with BK18, which are excluded from the baseline combination (semi-transparent points in the left panel of figure~\ref{fig:CMB_spectra}). The bounds on $\beta^2_{AF,T}$ and $\beta^2_{AF,S}$ loosen by only $\sim15\%$ with respect to the baseline combination, and remain tighter than those from SPA+BK18. With current data, the sensitivity to these parameters is therefore dominated by the $E$-to-$B$ leakage at $\ell\gtrsim300$ probed by SPT-3G D1, rather than by the $B$-mode measurements of BK18 at larger angular scales.

The bound on $\beta^2_F$, on the other hand, is essentially unaffected by which $B$-mode dataset is included, since in the absence of $V$-mode data this parameter is constrained through the rescaling of the $TE$ and $EE$ spectra by $\mathcal{Z}$ rather than through the $E$-to-$B$ leakage (see section~\ref{sec:current_data_results}). Nevertheless, without any $B$-mode data, the bound on $\beta^2_F$ is tighter. In this case all three LV parameters enter the spectra only through their weighted sum $\mathcal{Z}$, and, since their priors are positive, the larger values now allowed for the CPT-odd parameters leave less room for $\beta^2_F/4$.

\begin{table}[!t]
\centering
\renewcommand{\arraystretch}{1.4}
\resizebox{0.8\textwidth}{!}{%
\begin{tabular}{l c c c}
\hline\hline
\textbf{Dataset} & $\beta^2_{AF,T}\times10^4$ & $\beta^2_{AF,S}\times10^4$ & $\beta^2_{F}\times10^2$ \\
\hline
SPA+BK18+SPT$BB$+SPTpol$BB$ & $< 3.59$ & $< 3.65$ & $< 4.72$ \\
SPA+BK18+SPT$BB$            & $< 3.31$ & $< 3.31$ & $< 4.87$ \\
SPA+BK18+SPTpol$BB$         & $< 6.50$ & $< 6.52$ & $< 4.80$ \\
SPA+SPT$BB$+SPTpol$BB$         & $< 4.18$ & $< 4.23$ & $< 4.64$ \\
\hdashline
SPA                         & $< 79.8$ & $< 77.8$ & $< 3.84$ \\
\hline\hline
\end{tabular}
}
\caption{95\% CL upper limits on $\beta^2_{AF,T}$, $\beta^2_{AF,S}$ (in units of $10^{-4}$) and $\beta^2_{F}$ (in units of $10^{-2}$), for the baseline combination, for the combinations obtained by removing SPTpol, SPT-3G D1 or BK18 $B$-mode data individually, and for the SPA combination without any $B$-mode data.}
\label{tab:bmodes_individual_data}
\end{table}

\section{On the degeneracy with polarization calibration parameters}
\label{app:calibration}

\begin{figure}[!t]
    \centering
    \includegraphics[width=0.7\linewidth]{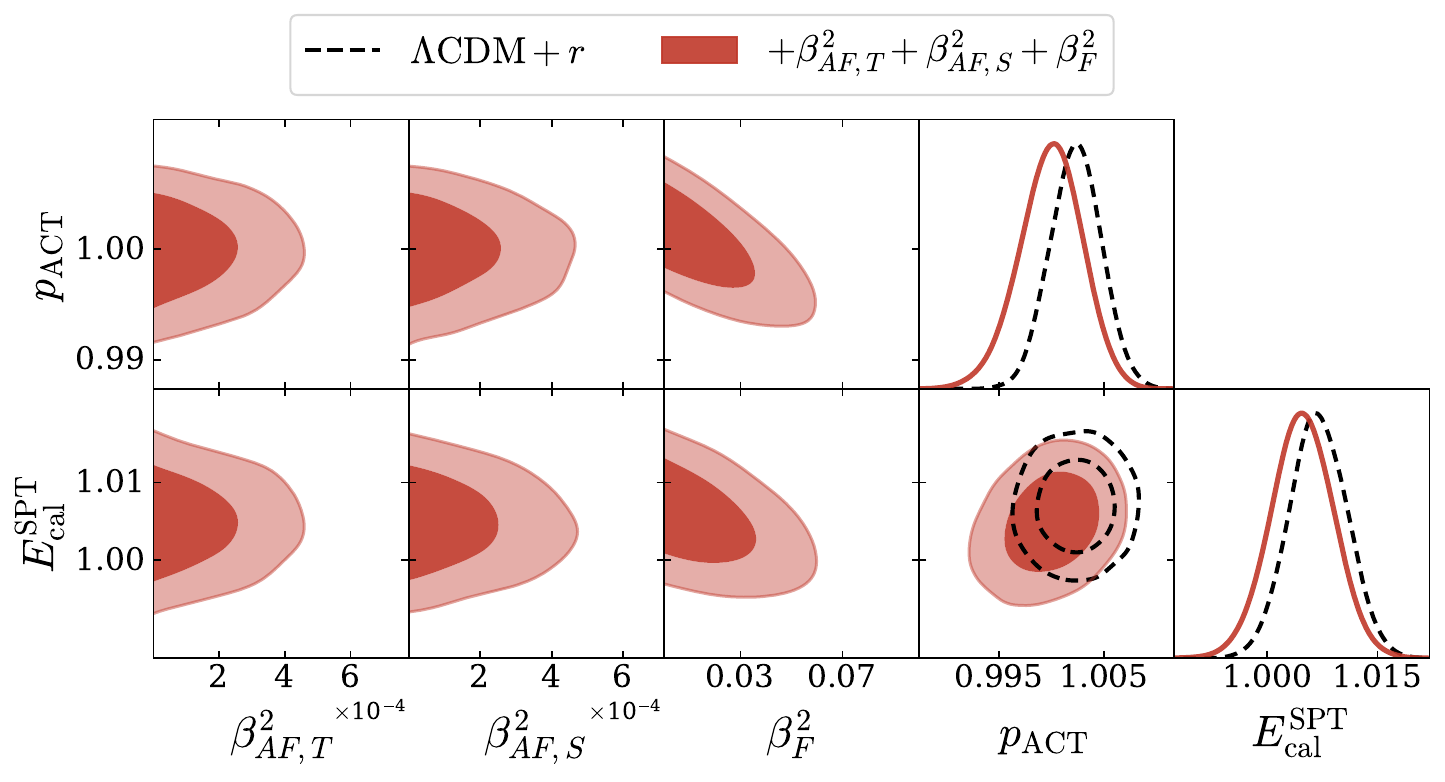}
    \caption{One- and two-dimensional posterior distributions of $\beta^2_{AF,T}$, $\beta^2_{AF,S}$ and $\beta^2_F$, together with the polarization calibration parameters of ACT and SPT ($p_{\rm ACT}$ and $E_{\rm cal}^{\rm SPT}$, respectively), for the baseline combination, SPA+BK18+SPT$BB$+SPTpol$BB$. For comparison, we also show the polarization calibration parameters obtained in the $\Lambda$CDM$+r$ analysis.}
    \label{fig:calibration_parameters}
\end{figure}

As discussed in section~\ref{sec:current_data_results}, in the absence of $V$-mode data $\beta^2_F$ enters the spectra only through the scale-independent rescaling of the $TE$ and $EE$ spectra by $\mathcal{Z}$ (eq.~\eqref{eq:Z}), which is partially degenerate with the polarization calibration parameters sampled in the ACT DR6 and SPT-3G D1 likelihoods, $p_{\rm ACT}$ and $E^{\rm SPT}_{\rm cal}$. In this appendix we examine this degeneracy in more detail and quantify its impact on the constraints.

\begin{figure}[!t]
    \centering
    \includegraphics[width=\linewidth]{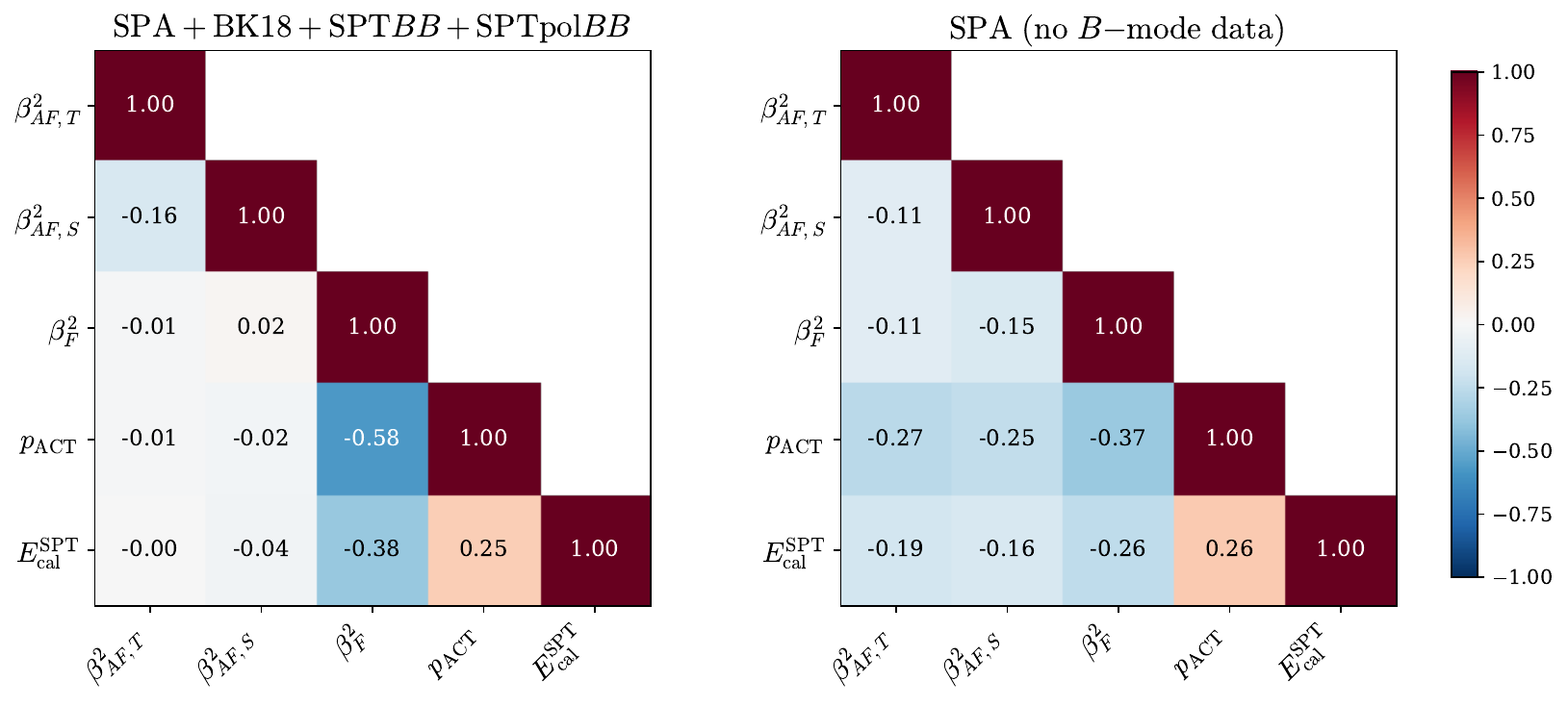}
    \caption{Correlation coefficients between the LV parameters and the polarization calibration parameters of ACT and SPT, $p_{\rm ACT}$ and $E^{\rm SPT}_{\rm cal}$, for the baseline combination (left) and for the SPA combination without any $B$-mode data (right).}
    \label{fig:calibration_correlations}
\end{figure}

Figure~\ref{fig:calibration_parameters} shows the posterior distributions of the LV parameters and of the two calibration parameters for the baseline combination, and figure~\ref{fig:calibration_correlations} (left) reports the corresponding correlation coefficients. $\beta^2_F$ is anticorrelated with both $p_{\rm ACT}$ and $E^{\rm SPT}_{\rm cal}$, with correlation coefficients of $-0.58$ and $-0.38$, respectively. Opening the LV parameter space also broadens the posteriors of both calibration parameters and shifts them towards lower values with respect to the $\Lambda$CDM$+r$ analysis, as expected if they compensate for a non-zero $\mathcal{Z}$. In contrast, the correlation between the calibration parameters and $\beta^2_{AF,T}$ or $\beta^2_{AF,S}$ is negligible. The latter also source a leakage of $E$ into $B$ modes, whose amplitude and scale dependence cannot be mimicked by a calibration of the polarization spectra, so that the $B$-mode data break the degeneracy. Removing all $B$-mode data confirms this, since for the SPA combination alone $\beta^2_{AF,T}$ and $\beta^2_{AF,S}$ become anticorrelated with both calibration parameters, like $\beta^2_F$ (figure~\ref{fig:calibration_correlations}, right panel).

\begin{figure}[!t]
    \centering
    \includegraphics[width=0.6\linewidth]{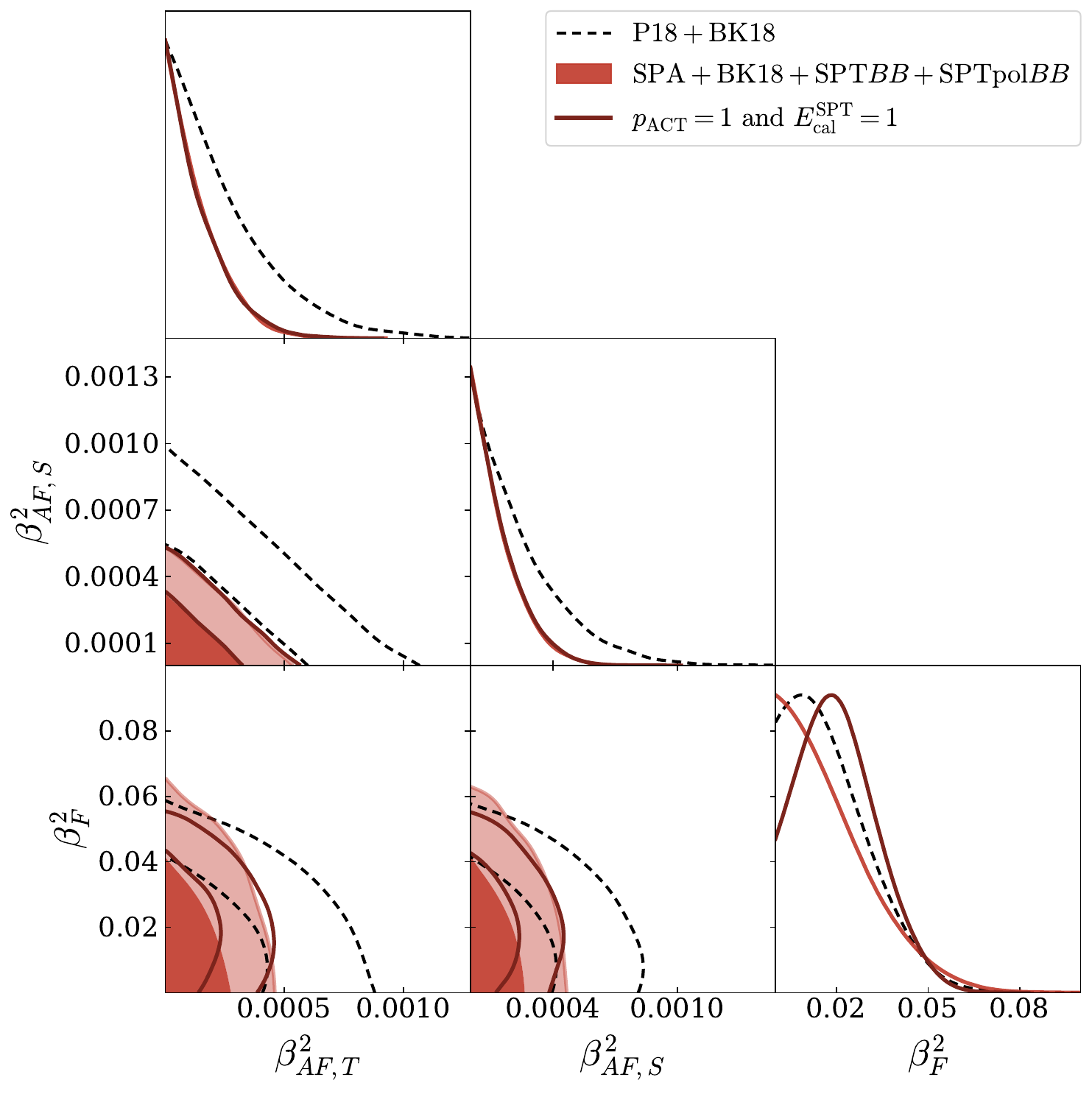}
    \caption{One- and two-dimensional posterior distributions of $\beta^2_{AF,T}$, $\beta^2_{AF,S}$ and $\beta^2_F$ for the baseline combination, with the polarization calibration parameters of ACT and SPT varied (red) or fixed to unity (dark red lines), compared with the P18+BK18 combination (black dashed).}
    \label{fig:calibration_fixed}
\end{figure}

To quantify the impact of this degeneracy, we repeat the analysis of the baseline combination fixing both calibration parameters to unity (figure~\ref{fig:calibration_fixed}). The bounds on $\beta^2_{AF,T}$ and $\beta^2_{AF,S}$ remain essentially unchanged, while the width of the $\beta^2_F$ posterior is reduced by about 15\%, showing that, once their calibration is fixed, ACT and SPT polarization data contribute to the constraint on $\mathcal{Z}$. The posterior also moves away from zero, with a mild preference for $\beta^2_F>0$ at 68\% CL, consistent with $E^{\rm SPT}_{\rm cal}$ favouring values slightly above unity in the $\Lambda$CDM$+r$ analysis (figure~\ref{fig:calibration_parameters}).

Similar considerations apply to \textit{Planck}, whose polarization efficiencies are not sampled in the \texttt{Plik\_lite} likelihood, but fixed to the values obtained by fitting the $EE$ spectrum of each frequency channel to the $\Lambda$CDM model that best fits $TT$~\cite{Planck:2019nip}. Any rescaling of the polarization spectra by $\mathcal{Z}$ would therefore have been absorbed into these efficiencies, and the \textit{Planck} bound on $\beta^2_F$, which dominates the constraint from current data, does not account for their uncertainty. As also noted in~\cite{AtacamaCosmologyTelescope:2025blo}, fixing the polarization efficiency can potentially disfavour extensions of $\Lambda$CDM that modify the amplitude of polarization relative to temperature.

Without $V$-mode data, the polarization calibration is therefore a key limitation for the constraint on $\beta^2_F$, which would benefit from an accurate external calibration of the polarization amplitude. Circular polarization can instead break this degeneracy, since the $VV$ spectrum sourced by the CPT-even operator cannot be mimicked by a miscalibration of the linear polarization. Current CLASS data already begin to do so, as including them in the baseline combination reduces the correlation coefficients of $\beta^2_F$ with $p_{\rm ACT}$ and $E^{\rm SPT}_{\rm cal}$ from $-0.58$ and $-0.38$ to $-0.49$ and $-0.34$, respectively.

\clearpage
\bibliographystyle{JHEP}
\bibliography{bibliography.bib}

@article{Pospelov:2008gg,
    author = "Pospelov, Maxim and Ritz, Adam and Skordis, Constantinos",
    title = "{Pseudoscalar perturbations and polarization of the cosmic microwave background}",
    eprint = "0808.0673",
    archivePrefix = "arXiv",
    primaryClass = "astro-ph",
    doi = "10.1103/PhysRevLett.103.051302",
    journal = "Phys. Rev. Lett.",
    volume = "103",
    pages = "051302",
    year = "2009"
}

@article{Carroll:1998zi,
    author = "Carroll, Sean M.",
    title = "{Quintessence and the rest of the world}",
    eprint = "astro-ph/9806099",
    archivePrefix = "arXiv",
    reportNumber = "NSF-ITP-98-063",
    doi = "10.1103/PhysRevLett.81.3067",
    journal = "Phys. Rev. Lett.",
    volume = "81",
    pages = "3067--3070",
    year = "1998"
}

@article{Carroll:1989vb,
    author = "Carroll, Sean M. and Field, George B. and Jackiw, Roman",
    title = "{Limits on a Lorentz and Parity Violating Modification of Electrodynamics}",
    reportNumber = "MIT-CTP-1782",
    doi = "10.1103/PhysRevD.41.1231",
    journal = "Phys. Rev. D",
    volume = "41",
    pages = "1231",
    year = "1990"
}

@article{Li:2008tma,
    author = "Li, Mingzhe and Zhang, Xinmin",
    title = "{Cosmological CPT violating effect on CMB polarization}",
    eprint = "0810.0403",
    archivePrefix = "arXiv",
    primaryClass = "astro-ph",
    doi = "10.1103/PhysRevD.78.103516",
    journal = "Phys. Rev. D",
    volume = "78",
    pages = "103516",
    year = "2008"
}

@article{Addazi:2021xuf,
    author = "Addazi, A. and others",
    title = "{Quantum gravity phenomenology at the dawn of the multi-messenger era\textemdash{}A review}",
    eprint = "2111.05659",
    archivePrefix = "arXiv",
    primaryClass = "hep-ph",
    doi = "10.1016/j.ppnp.2022.103948",
    journal = "Prog. Part. Nucl. Phys.",
    volume = "125",
    pages = "103948",
    year = "2022"
}

@article{Amelino-Camelia:2008aez,
    author = "Amelino-Camelia, Giovanni",
    title = "{Quantum-Spacetime Phenomenology}",
    eprint = "0806.0339",
    archivePrefix = "arXiv",
    primaryClass = "gr-qc",
    doi = "10.12942/lrr-2013-5",
    journal = "Living Rev. Rel.",
    volume = "16",
    pages = "5",
    year = "2013"
}

@article{Mattingly:2005re,
    author = "Mattingly, David",
    title = "{Modern tests of Lorentz invariance}",
    eprint = "gr-qc/0502097",
    archivePrefix = "arXiv",
    doi = "10.12942/lrr-2005-5",
    journal = "Living Rev. Rel.",
    volume = "8",
    pages = "5",
    year = "2005"
}

@article{Torrado:2020dgo,
    author = "Torrado, Jesus and Lewis, Antony",
    title = "{Cobaya: Code for Bayesian Analysis of hierarchical physical models}",
    eprint = "2005.05290",
    archivePrefix = "arXiv",
    primaryClass = "astro-ph.IM",
    reportNumber = "TTK-20-15",
    doi = "10.1088/1475-7516/2021/05/057",
    journal = "JCAP",
    volume = "05",
    pages = "057",
    year = "2021"
}

@article{Howlett:2012mh,
    author = "Howlett, Cullan and Lewis, Antony and Hall, Alex and Challinor, Anthony",
    title = "{CMB power spectrum parameter degeneracies in the era of precision cosmology}",
    eprint = "1201.3654",
    archivePrefix = "arXiv",
    primaryClass = "astro-ph.CO",
    doi = "10.1088/1475-7516/2012/04/027",
    journal = "JCAP",
    volume = "04",
    pages = "027",
    year = "2012"
}

@article{Padilla:2019dhz,
doi = {10.3847/1538-4357/ab61f8},
url = {https://dx.doi.org/10.3847/1538-4357/ab61f8},
year = {2020},
month = {jan},
publisher = {The American Astronomical Society},
volume = {889},
number = {2},
pages = {105},
author = {Ivan L. Padilla and Joseph R. Eimer and Yunyang Li and Graeme E. Addison and Aamir Ali and John W. Appel and Charles L. Bennett and Ricardo Bustos and Michael K. Brewer and Manwei Chan and David T. Chuss and Joseph Cleary and Jullianna Couto and Sumit Dahal and Kevin Denis and Rolando Dünner and Thomas Essinger-Hileman and Pedro Fluxá and Dominik Gothe and Saianeesh K. Haridas and Kathleen Harrington and Jeffrey Iuliano and John Karakla and Tobias A. Marriage and Nathan J. Miller and Carolina Núñez and Lucas Parker and Matthew A. Petroff and Rodrigo Reeves and Karwan Rostem and Robert W. Stevens and Deniz Augusto Nunes Valle and Duncan J. Watts and Janet L. Weiland and Edward J. Wollack and Zhilei Xu},
title = {Two-year Cosmology Large Angular Scale Surveyor (CLASS) Observations: A Measurement of Circular Polarization at 40 GHz},
journal = "Astrophys. J.",
eprint = "1911.00391",
archivePrefix = "arXiv",
primaryClass = "astro-ph.CO"
}

@article{SPIDER:2017kwu,
    author = "Nagy, J. M. and others",
    collaboration = "SPIDER",
    title = "{A New Limit on CMB Circular Polarization from SPIDER}",
    eprint = "1704.00215",
    archivePrefix = "arXiv",
    primaryClass = "astro-ph.CO",
    reportNumber = "FERMILAB-PUB-17-135-AE",
    doi = "10.3847/1538-4357/aa7cfd",
    journal = "Astrophys. J.",
    volume = "844",
    number = "2",
    pages = "151",
    year = "2017"
}

@article{Kostelecky:2008ts,
    author = "Kostelecky, V. Alan and Russell, Neil",
    title = "{Data Tables for Lorentz and CPT Violation}",
    eprint = "0801.0287",
    archivePrefix = "arXiv",
    primaryClass = "hep-ph",
    month = "1",
    year = "2008"
}

@article{Lembo:2020ufn,
    author = "Lembo, Margherita and Lattanzi, Massimiliano and Pagano, Luca and Gruppuso, Alessandro and Natoli, Paolo and Forastieri, Francesco",
    title = "{Cosmic Microwave Background Polarization as a Tool to Constrain the Optical Properties of the Universe}",
    eprint = "2007.08486",
    archivePrefix = "arXiv",
    primaryClass = "astro-ph.CO",
    doi = "10.1103/PhysRevLett.127.011301",
    journal = "Phys. Rev. Lett.",
    volume = "127",
    number = "1",
    pages = "011301",
    year = "2021"
}

@article{Gubitosi:2012rg,
    author = "Gubitosi, G. and Paci, F.",
    title = "{Constraints on cosmological birefringence energy dependence from CMB polarization data}",
    eprint = "1211.3321",
    archivePrefix = "arXiv",
    primaryClass = "astro-ph.CO",
    doi = "10.1088/1475-7516/2013/02/020",
    journal = "JCAP",
    volume = "02",
    pages = "020",
    year = "2013"
}

@article{BICEP:2021xfz,
    author = "Ade, P. A. R. and others",
    collaboration = "BICEP, Keck",
    title = "{Improved Constraints on Primordial Gravitational Waves using Planck, WMAP, and BICEP/Keck Observations through the 2018 Observing Season}",
    eprint = "2110.00483",
    archivePrefix = "arXiv",
    primaryClass = "astro-ph.CO",
    doi = "10.1103/PhysRevLett.127.151301",
    journal = "Phys. Rev. Lett.",
    volume = "127",
    number = "15",
    pages = "151301",
    year = "2021"
}

@article{Gelman:1992zz,
    author = "Gelman, Andrew and Rubin, Donald B.",
    title = "{Inference from Iterative Simulation Using Multiple Sequences}",
    doi = "10.1214/ss/1177011136",
    journal = "Statist. Sci.",
    volume = "7",
    pages = "457--472",
    year = "1992"
}

@article{ACT:2020gnv,
    author = "Aiola, Simone and others",
    collaboration = "ACT",
    title = "{The Atacama Cosmology Telescope: DR4 Maps and Cosmological Parameters}",
    eprint = "2007.07288",
    archivePrefix = "arXiv",
    primaryClass = "astro-ph.CO",
    doi = "10.1088/1475-7516/2020/12/047",
    journal = "JCAP",
    volume = "12",
    pages = "047",
    year = "2020"
}

@article{Gerasimov:2021chj,
    author = "Gerasimov, Roman and Bhoj, Praneet and Kislat, Fabian",
    title = "{New Constraints on Lorentz Invariance Violation from Combined Linear and Circular Optical Polarimetry of Extragalactic Sources}",
    eprint = "2104.00238",
    archivePrefix = "arXiv",
    primaryClass = "astro-ph.CO",
    doi = "10.3390/sym13050880",
    journal = "Symmetry",
    volume = "13",
    number = "5",
    pages = "880",
    year = "2021"
}

@article{Friedman:2020bxa,
    author = "Friedman, Andrew S. and Gerasimov, Roman and Leon, David and Stevens, Walker and Tytler, David and Keating, Brian G. and Kislat, Fabian",
    title = "{Improved constraints on anisotropic birefringent Lorentz invariance and $CPT$ violation from broadband optical polarimetry of high redshift galaxies}",
    eprint = "2003.00647",
    archivePrefix = "arXiv",
    primaryClass = "astro-ph.HE",
    doi = "10.1103/PhysRevD.102.043008",
    journal = "Phys. Rev. D",
    volume = "102",
    number = "4",
    pages = "043008",
    year = "2020"
}

@article{Lewis:2019xzd,
 author         = "Lewis, Antony",
 title          = "{GetDist: a Python package for analysing Monte Carlo
                   samples}",
 year           = "2019",
 eprint         = "1910.13970",
 archivePrefix  = "arXiv",
 primaryClass   = "astro-ph.IM",
 SLACcitation   = "%%CITATION = ARXIV:1910.13970;%%",
 url            = "https://getdist.readthedocs.io"
}

@article{Harris:2020xlr,
    author = "Harris, Charles R. and others",
    title = "{Array programming with NumPy}",
    eprint = "2006.10256",
    archivePrefix = "arXiv",
    primaryClass = "cs.MS",
    doi = "10.1038/s41586-020-2649-2",
    journal = "Nature",
    volume = "585",
    number = "7825",
    pages = "357--362",
    year = "2020"
}

@article{Hunter:2007ouj,
    author = "Hunter, John D.",
    title = "{Matplotlib: A 2D Graphics Environment}",
    doi = "10.1109/MCSE.2007.55",
    journal = "Comput. Sci. Eng.",
    volume = "9",
    number = "3",
    pages = "90--95",
    year = "2007"
}

@article{Minami:2019ruj,
    author = "Minami, Yuto and Ochi, Hiroki and Ichiki, Kiyotomo and Katayama, Nobuhiko and Komatsu, Eiichiro and Matsumura, Tomotake",
    title = "{Simultaneous determination of the cosmic birefringence and miscalibrated polarization angles from CMB experiments}",
    eprint = "1904.12440",
    archivePrefix = "arXiv",
    primaryClass = "astro-ph.CO",
    doi = "10.1093/ptep/ptz079",
    journal = "PTEP",
    volume = "2019",
    number = "8",
    pages = "083E02",
    year = "2019"
}

@article{Clark:2021kze,
    author = "Clark, S. E. and Kim, Chang-Goo and Hill, J. Colin and Hensley, Brandon S.",
    title = "{The Origin of Parity Violation in Polarized Dust Emission and Implications for Cosmic Birefringence}",
    eprint = "2105.00120",
    archivePrefix = "arXiv",
    primaryClass = "astro-ph.GA",
    doi = "10.3847/1538-4357/ac0e35",
    journal = "Astrophys. J.",
    volume = "919",
    number = "1",
    pages = "53",
    year = "2021"
}

@article{Minami:2020xfg,
    author = "Minami, Yuto",
    title = "{Determination of miscalibrated polarization angles from observed cosmic microwave background and foreground $EB$ power spectra: Application to partial-sky observation}",
    eprint = "2002.03572",
    archivePrefix = "arXiv",
    primaryClass = "astro-ph.CO",
    doi = "10.1093/ptep/ptaa057",
    journal = "PTEP",
    volume = "2020",
    number = "6",
    pages = "063E01",
    year = "2020"
}

@article{SimonsObservatory:2018koc,
    author = "Ade, Peter and others",
    collaboration = "Simons Observatory",
    title = "{The Simons Observatory: Science goals and forecasts}",
    eprint = "1808.07445",
    archivePrefix = "arXiv",
    primaryClass = "astro-ph.CO",
    doi = "10.1088/1475-7516/2019/02/056",
    journal = "JCAP",
    volume = "02",
    pages = "056",
    year = "2019"
}

@article{Minami:2020odp,
    author = "Minami, Yuto and Komatsu, Eiichiro",
    title = "{New Extraction of the Cosmic Birefringence from the Planck 2018 Polarization Data}",
    eprint = "2011.11254",
    archivePrefix = "arXiv",
    primaryClass = "astro-ph.CO",
    doi = "10.1103/PhysRevLett.125.221301",
    journal = "Phys. Rev. Lett.",
    volume = "125",
    number = "22",
    pages = "221301",
    year = "2020"
}

@article{Lue:1998mq,
    author = "Lue, Arthur and Wang, Li-Min and Kamionkowski, Marc",
    title = "{Cosmological signature of new parity violating interactions}",
    eprint = "astro-ph/9812088",
    archivePrefix = "arXiv",
    reportNumber = "CU-TP-926, CAL-675",
    doi = "10.1103/PhysRevLett.83.1506",
    journal = "Phys. Rev. Lett.",
    volume = "83",
    pages = "1506--1509",
    year = "1999"
}

@article{Hu:2001bc,
    author = "Hu, Wayne and Dodelson, Scott",
    title = "{Cosmic Microwave Background Anisotropies}",
    eprint = "astro-ph/0110414",
    archivePrefix = "arXiv",
    reportNumber = "FERMILAB-PUB-01-328-A",
    doi = "10.1146/annurev.astro.40.060401.093926",
    journal = "Ann. Rev. Astron. Astrophys.",
    volume = "40",
    pages = "171--216",
    year = "2002"
}

@article{King:2016exc,
    author = "King, Soma and Lubin, Philip",
    title = "{Circular polarization of the CMB: Foregrounds and detection prospects}",
    eprint = "1606.04112",
    archivePrefix = "arXiv",
    primaryClass = "astro-ph.CO",
    doi = "10.1103/PhysRevD.94.023501",
    journal = "Phys. Rev. D",
    volume = "94",
    number = "2",
    pages = "023501",
    year = "2016"
}

@article{Montero-Camacho:2018vgs,
    author = "Montero-Camacho, Paulo and Hirata, Christopher M.",
    title = "{Exploring circular polarization in the CMB due to conventional sources of cosmic birefringence}",
    eprint = "1803.04505",
    archivePrefix = "arXiv",
    primaryClass = "astro-ph.CO",
    doi = "10.1088/1475-7516/2018/08/040",
    journal = "JCAP",
    volume = "08",
    pages = "040",
    year = "2018"
}

@article{Alexander:2008fp,
    author = "Alexander, Stephon and Ochoa, Joseph and Kosowsky, Arthur",
    title = "{Generation of Circular Polarization of the Cosmic Microwave Background}",
    eprint = "0810.2355",
    archivePrefix = "arXiv",
    primaryClass = "astro-ph",
    doi = "10.1103/PhysRevD.79.063524",
    journal = "Phys. Rev. D",
    volume = "79",
    pages = "063524",
    year = "2009"
}

@article{Inomata:2018vbu,
    author = "Inomata, Keisuke and Kamionkowski, Marc",
    title = "{Circular polarization of the cosmic microwave background from vector and tensor perturbations}",
    eprint = "1811.04957",
    archivePrefix = "arXiv",
    primaryClass = "astro-ph.CO",
    reportNumber = "IPMU 18-0184",
    doi = "10.1103/PhysRevD.99.043501",
    journal = "Phys. Rev. D",
    volume = "99",
    number = "4",
    pages = "043501",
    year = "2019"
}

@article{Ejlli:2016avx,
    author = "Ejlli, Damian",
    title = "{Magneto-optic effects of the cosmic microwave background}",
    eprint = "1607.02094",
    archivePrefix = "arXiv",
    primaryClass = "astro-ph.CO",
    doi = "10.1016/j.nuclphysb.2018.08.003",
    journal = "Nucl. Phys. B",
    volume = "935",
    pages = "83--128",
    year = "2018"
}

@article{Bartolo:2019eac,
    author = "Bartolo, Nicola and Hoseinpour, Ahmad and Matarrese, Sabino and Orlando, Giorgio and Zarei, Moslem",
    title = "{CMB Circular and B-mode Polarization from New Interactions}",
    eprint = "1903.04578",
    archivePrefix = "arXiv",
    primaryClass = "hep-ph",
    doi = "10.1103/PhysRevD.100.043516",
    journal = "Phys. Rev. D",
    volume = "100",
    number = "4",
    pages = "043516",
    year = "2019"
}

@article{Ejlli:2017uli,
    author = "Ejlli, Damian",
    title = "{Millicharged fermion vacuum polarization in a cosmic magnetic field and generation of CMB elliptic polarization}",
    eprint = "1704.01894",
    archivePrefix = "arXiv",
    primaryClass = "hep-ph",
    doi = "10.1103/PhysRevD.96.023540",
    journal = "Phys. Rev. D",
    volume = "96",
    number = "2",
    pages = "023540",
    year = "2017"
}

@article{Tizchang:2016vef,
    author = "Tizchang, S. and Batebi, S. and Haghighat, M. and Mohammadi, R.",
    title = "{Cosmic microwave background polarization in non-commutative space-time}",
    eprint = "1605.09045",
    archivePrefix = "arXiv",
    primaryClass = "hep-ph",
    doi = "10.1140/epjc/s10052-016-4312-5",
    journal = "Eur. Phys. J. C",
    volume = "76",
    number = "9",
    pages = "478",
    year = "2016"
}

@article{Bavarsad:2009hm,
    author = "Bavarsad, E. and Haghighat, M. and Rezaei, Z. and Mohammadi, R. and Motie, I. and Zarei, M.",
    title = "{Generation of circular polarization of the CMB}",
    eprint = "0912.2993",
    archivePrefix = "arXiv",
    primaryClass = "hep-th",
    doi = "10.1103/PhysRevD.81.084035",
    journal = "Phys. Rev. D",
    volume = "81",
    pages = "084035",
    year = "2010"
}

@inproceedings{LiteBIRD:2026gdf,
    author = "Aizawa, K. and others",
    collaboration = "LiteBIRD",
    title = "{LiteBIRD Mission Overview after Mission Reformation}",
    booktitle = "{SPIE Astronomical Telescopes + Instrumentation 2026}",
    eprint = "2607.25285",
    archivePrefix = "arXiv",
    primaryClass = "astro-ph.IM",
    month = "7",
    year = "2026"
}

@article{Diego-Palazuelos:2022dsq,
    author = "Diego-Palazuelos, P. and others",
    title = "{Cosmic Birefringence from the Planck Data Release 4}",
    eprint = "2201.07682",
    archivePrefix = "arXiv",
    primaryClass = "astro-ph.CO",
    doi = "10.1103/PhysRevLett.128.091302",
    journal = "Phys. Rev. Lett.",
    volume = "128",
    number = "9",
    pages = "091302",
    year = "2022"
}

@article{Eskilt:2022cff,
	author = "Eskilt, Johannes R. and Komatsu, Eiichiro",
	title = "{Improved constraints on cosmic birefringence from the WMAP and Planck cosmic microwave background polarization data}",
	eprint = "2205.13962",
	archivePrefix = "arXiv",
	primaryClass = "astro-ph.CO",
	doi = "10.1103/PhysRevD.106.063503",
	journal = "Phys. Rev. D",
	volume = "106",
	number = "6",
	pages = "063503",
	year = "2022"
}

@article{Jain:2021shf,
	author = "Jain, Mudit and Long, Andrew J. and Amin, Mustafa A.",
	title = "{CMB birefringence from ultralight-axion string networks}",
	eprint = "2103.10962",
	archivePrefix = "arXiv",
	primaryClass = "astro-ph.CO",
	doi = "10.1088/1475-7516/2021/05/055",
	journal = "JCAP",
	volume = "05",
	pages = "055",
	year = "2021"
}

@article{Ferreira:2023jbu,
	author = "Ferreira, Ricardo Z. and Gasparotto, Silvia and Hiramatsu, Takashi and Obata, Ippei and Pujolas, Oriol",
	title = "{Axionic defects in the CMB: birefringence and gravitational waves}",
	eprint = "2312.14104",
	archivePrefix = "arXiv",
	primaryClass = "hep-ph",
	reportNumber = "RUP-23-27",
	doi = "10.1088/1475-7516/2024/05/066",
	journal = "JCAP",
	volume = "05",
	pages = "066",
	year = "2024"
}

@article{Mangano:2001iu,
	author = "Mangano, G. and Miele, G. and Pastor, S. and Peloso, M.",
	title = "{A Precision calculation of the effective number of cosmological neutrinos}",
	eprint = "astro-ph/0111408",
	archivePrefix = "arXiv",
	reportNumber = "DSF-37-2001, MPI-PHT-2001-51",
	doi = "10.1016/S0370-2693(02)01622-2",
	journal = "Phys. Lett. B",
	volume = "534",
	pages = "8--16",
	year = "2002"
}

@article{Bennett:2019ewm,
	author = "Bennett, Jack J. and Buldgen, Gilles and Drewes, Marco and Wong, Yvonne Y. Y.",
	title = "{Towards a precision calculation of the effective number of neutrinos $N_{\rm eff}$ in the Standard Model I: the QED equation of state}",
	eprint = "1911.04504",
	archivePrefix = "arXiv",
	primaryClass = "hep-ph",
	doi = "10.1088/1475-7516/2020/03/003",
	journal = "JCAP",
	volume = "03",
	pages = "003",
	year = "2020",
	note = "[Addendum: JCAP 03, A01 (2021)]"
}

@article{Bennett:2020zkv,
	author = "Bennett, Jack J. and Buldgen, Gilles and De Salas, Pablo F. and Drewes, Marco and Gariazzo, Stefano and Pastor, Sergio and Wong, Yvonne Y. Y.",
	title = "{Towards a precision calculation of $N_{\rm eff}$ in the Standard Model II: Neutrino decoupling in the presence of flavour oscillations and finite-temperature QED}",
	eprint = "2012.02726",
	archivePrefix = "arXiv",
	primaryClass = "hep-ph",
	reportNumber = "CPPC-2020-10",
	doi = "10.1088/1475-7516/2021/04/073",
	journal = "JCAP",
	volume = "04",
	pages = "073",
	year = "2021"
}

@article{Akita:2020szl,
	author = "Akita, Kensuke and Yamaguchi, Masahide",
	title = "{A precision calculation of relic neutrino decoupling}",
	eprint = "2005.07047",
	archivePrefix = "arXiv",
	primaryClass = "hep-ph",
	doi = "10.1088/1475-7516/2020/08/012",
	journal = "JCAP",
	volume = "08",
	pages = "012",
	year = "2020"
}

@article{Froustey:2020mcq,
	author = "Froustey, Julien and Pitrou, Cyril and Volpe, Maria Cristina",
	title = "{Neutrino decoupling including flavour oscillations and primordial nucleosynthesis}",
	eprint = "2008.01074",
	archivePrefix = "arXiv",
	primaryClass = "hep-ph",
	doi = "10.1088/1475-7516/2020/12/015",
	journal = "JCAP",
	volume = "12",
	pages = "015",
	year = "2020"
}

@article{Planck:2018nkj,
	author = "Aghanim, N. and others",
	collaboration = "Planck",
	title = "{Planck 2018 results. I. Overview and the cosmological legacy of Planck}",
	eprint = "1807.06205",
	archivePrefix = "arXiv",
	primaryClass = "astro-ph.CO",
	doi = "10.1051/0004-6361/201833880",
	journal = "Astron. Astrophys.",
	volume = "641",
	pages = "A1",
	year = "2020"
}

@article{Planck:2018vyg,
	author = "Aghanim, N. and others",
	collaboration = "Planck",
	title = "{Planck 2018 results. VI. Cosmological parameters}",
	eprint = "1807.06209",
	archivePrefix = "arXiv",
	primaryClass = "astro-ph.CO",
	doi = "10.1051/0004-6361/201833910",
	journal = "Astron. Astrophys.",
	volume = "641",
	pages = "A6",
	year = "2020",
	note = "[Erratum: Astron.Astrophys. 652, C4 (2021)]"
}

@article{Drewes:2024wbw,
    author = "Drewes, Marco and Georis, Yannis and Klasen, Michael and Wiggering, Luca Paolo and Wong, Yvonne Y. Y.",
    title = "{Towards a precision calculation of N $_{eff}$ in the Standard Model. Part III. Improved estimate of NLO contributions to the collision integral}",
    eprint = "2402.18481",
    archivePrefix = "arXiv",
    primaryClass = "hep-ph",
    reportNumber = "CPPC-2024-01, MS-TP-24-06",
    doi = "10.1088/1475-7516/2024/06/032",
    journal = "JCAP",
    volume = "06",
    pages = "032",
    year = "2024"
}

@article{LiteBIRD:2022cnt,
    author = "Allys, E. and others",
    collaboration = "LiteBIRD",
    title = "{Probing Cosmic Inflation with the LiteBIRD Cosmic Microwave Background Polarization Survey}",
    eprint = "2202.02773",
    archivePrefix = "arXiv",
    primaryClass = "astro-ph.IM",
    doi = "10.1093/ptep/ptac150",
    journal = "PTEP",
    volume = "2023",
    number = "4",
    pages = "042F01",
    year = "2023"
}

@article{Caloni:2022kwp,
    author = "Caloni, Luca and Giardiello, Serena and Lembo, Margherita and Gerbino, Martina and Gubitosi, Giulia and Lattanzi, Massimiliano and Pagano, Luca",
    title = "{Probing Lorentz-violating electrodynamics with CMB polarization}",
    eprint = "2212.04867",
    archivePrefix = "arXiv",
    primaryClass = "astro-ph.CO",
    doi = "10.1088/1475-7516/2023/03/018",
    journal = "JCAP",
    volume = "03",
    pages = "018",
    year = "2023"
}

@article{Colladay:1998fq,
	author = "Colladay, Don and Kosteleck{\'y}, V. Alan",
	title = "{Lorentz violating extension of the standard model}",
	eprint = "hep-ph/9809521",
	archivePrefix = "arXiv",
	reportNumber = "IUHET-359",
	doi = "10.1103/PhysRevD.58.116002",
	journal = "Phys. Rev. D",
	volume = "58",
	pages = "116002",
	year = "1998"
}

@article{Kostelecky:2007zz,
	author = "Kosteleck{\'y}, V. Alan and Mewes, Matthew",
	title = "{Lorentz-violating electrodynamics and the cosmic microwave background}",
	eprint = "astro-ph/0702379",
	archivePrefix = "arXiv",
	reportNumber = "IUHET-504",
	doi = "10.1103/PhysRevLett.99.011601",
	journal = "Phys. Rev. Lett.",
	volume = "99",
	pages = "011601",
	year = "2007"
}

@article{Planck:2016soo,
    author = "Aghanim, N. and others",
    collaboration = "Planck",
    title = "{Planck intermediate results. XLIX. Parity-violation constraints from polarization data}",
    eprint = "1605.08633",
    archivePrefix = "arXiv",
    primaryClass = "astro-ph.CO",
    doi = "10.1051/0004-6361/201629018",
    journal = "Astron. Astrophys.",
    volume = "596",
    pages = "A110",
    year = "2016"
}

@article{Gluscevic:2009mm,
    author = "Gluscevic, Vera and Kamionkowski, Marc and Cooray, Asantha",
    title = "{De-Rotation of the Cosmic Microwave Background Polarization: Full-Sky Formalism}",
    eprint = "0905.1687",
    archivePrefix = "arXiv",
    primaryClass = "astro-ph.CO",
    doi = "10.1103/PhysRevD.80.023510",
    journal = "Phys. Rev. D",
    volume = "80",
    pages = "023510",
    year = "2009"
}

@article{Zagatti:2024jxm,
    author = "Zagatti, G. and Bortolami, M. and Gruppuso, A. and Natoli, P. and Pagano, L. and Fabbian, G.",
    title = "{Planck constraints on cosmic birefringence and its cross-correlation with the CMB}",
    eprint = "2401.11973",
    archivePrefix = "arXiv",
    primaryClass = "astro-ph.CO",
    doi = "10.1088/1475-7516/2024/05/034",
    journal = "JCAP",
    volume = "05",
    pages = "034",
    year = "2024"
}

@article{Lewis:1999bs,
    author = "Lewis, Antony and Challinor, Anthony and Lasenby, Anthony",
    title = "{Efficient computation of CMB anisotropies in closed FRW models}",
    eprint = "astro-ph/9911177",
    archivePrefix = "arXiv",
    doi = "10.1086/309179",
    journal = "Astrophys. J.",
    volume = "538",
    pages = "473--476",
    year = "2000"
}

@article{Planck:2019nip,
    author = "Aghanim, N. and others",
    collaboration = "Planck",
    title = "{Planck 2018 results. V. CMB power spectra and likelihoods}",
    eprint = "1907.12875",
    archivePrefix = "arXiv",
    primaryClass = "astro-ph.CO",
    doi = "10.1051/0004-6361/201936386",
    journal = "Astron. Astrophys.",
    volume = "641",
    pages = "A5",
    year = "2020"
}

@article{Pagano:2019tci,
    author = "Pagano, L. and Delouis, J. -M. and Mottet, S. and Puget, J. -L. and Vibert, L.",
    title = "{Reionization optical depth determination from Planck HFI data with ten percent accuracy}",
    eprint = "1908.09856",
    archivePrefix = "arXiv",
    primaryClass = "astro-ph.CO",
    doi = "10.1051/0004-6361/201936630",
    journal = "Astron. Astrophys.",
    volume = "635",
    pages = "A99",
    year = "2020"
}

@article{Delouis:2019bub,
    author = "Delouis, J. -M. and Pagano, L. and Mottet, S. and Puget, J. -L. and Vibert, L.",
    title = "{SRoll2: an improved mapmaking approach to reduce large-scale systematic effects in the Planck High Frequency Instrument legacy maps}",
    eprint = "1901.11386",
    archivePrefix = "arXiv",
    primaryClass = "astro-ph.CO",
    doi = "10.1051/0004-6361/201834882",
    journal = "Astron. Astrophys.",
    volume = "629",
    pages = "A38",
    year = "2019"
}

@article{Carron:2022eyg,
    author = "Carron, Julien and Mirmelstein, Mark and Lewis, Antony",
    title = "{CMB lensing from Planck PR4~maps}",
    eprint = "2206.07773",
    archivePrefix = "arXiv",
    primaryClass = "astro-ph.CO",
    doi = "10.1088/1475-7516/2022/09/039",
    journal = "JCAP",
    volume = "09",
    pages = "039",
    year = "2022"
}

@article{Perotto:2006rj,
    author = "Perotto, Laurence and Lesgourgues, Julien and Hannestad, Steen and Tu, Huitzu and Wong, Yvonne Y. Y.",
    title = "{Probing cosmological parameters with the CMB: Forecasts from full Monte Carlo simulations}",
    eprint = "astro-ph/0606227",
    archivePrefix = "arXiv",
    reportNumber = "MPP-2006-76, LAPTH-1148-06",
    doi = "10.1088/1475-7516/2006/10/013",
    journal = "JCAP",
    volume = "10",
    pages = "013",
    year = "2006"
}

@article{Wu:2014hta,
    author = "Wu, W. L. K. and Errard, J. and Dvorkin, C. and Kuo, C. L. and Lee, A. T. and McDonald, P. and Slosar, A. and Zahn, O.",
    title = "{A Guide to Designing Future Ground-based Cosmic Microwave Background Experiments}",
    eprint = "1402.4108",
    archivePrefix = "arXiv",
    primaryClass = "astro-ph.CO",
    doi = "10.1088/0004-637X/788/2/138",
    journal = "Astrophys. J.",
    volume = "788",
    pages = "138",
    year = "2014"
}

@article{Gerbino:2019okg,
    author = "Gerbino, Martina and Lattanzi, Massimiliano and Migliaccio, Marina and Pagano, Luca and Salvati, Laura and Colombo, Loris and Gruppuso, Alessandro and Natoli, Paolo and Polenta, Gianluca",
    title = "{Likelihood methods for CMB experiments}",
    eprint = "1909.09375",
    archivePrefix = "arXiv",
    primaryClass = "astro-ph.CO",
    doi = "10.3389/fphy.2020.00015",
    journal = "Front. in Phys.",
    volume = "8",
    pages = "15",
    year = "2020"
}

@article{Raffuzzi:2024wyh,
    author = "Raffuzzi, N. and Lembo, M. and Giardiello, S. and Gerbino, M. and Lattanzi, M. and Natoli, P. and Pagano, L.",
    title = "{Unveiling V modes: enhancing CMB sensitivity to BSM physics with a non-ideal half-wave plate}",
    eprint = "2409.10424",
    archivePrefix = "arXiv",
    primaryClass = "astro-ph.CO",
    doi = "10.1088/1475-7516/2025/03/009",
    journal = "JCAP",
    volume = "03",
    pages = "009",
    year = "2025"
}

@article{deSalas:2020pgw,
    author = "de Salas, P. F. and Forero, D. V. and Gariazzo, S. and Mart{\'\i}nez-Mirav{\'e}, P. and Mena, O. and Ternes, C. A. and T{\'o}rtola, M. and Valle, J. W. F.",
    title = "{2020 global reassessment of the neutrino oscillation picture}",
    eprint = "2006.11237",
    archivePrefix = "arXiv",
    primaryClass = "hep-ph",
    doi = "10.1007/JHEP02(2021)071",
    journal = "JHEP",
    volume = "02",
    pages = "071",
    year = "2021"
}

@article{Capozzi:2021fjo,
    author = "Capozzi, Francesco and Di Valentino, Eleonora and Lisi, Eligio and Marrone, Antonio and Melchiorri, Alessandro and Palazzo, Antonio",
    title = "{Unfinished fabric of the three neutrino paradigm}",
    eprint = "2107.00532",
    archivePrefix = "arXiv",
    primaryClass = "hep-ph",
    doi = "10.1103/PhysRevD.104.083031",
    journal = "Phys. Rev. D",
    volume = "104",
    number = "8",
    pages = "083031",
    year = "2021"
}

@article{Esteban:2020cvm,
    author = "Esteban, Ivan and Gonzalez-Garcia, M. C. and Maltoni, Michele and Schwetz, Thomas and Zhou, Albert",
    title = "{The fate of hints: updated global analysis of three-flavor neutrino oscillations}",
    eprint = "2007.14792",
    archivePrefix = "arXiv",
    primaryClass = "hep-ph",
    reportNumber = "IFT-UAM/CSIC-112, YITP-SB-2020-21",
    doi = "10.1007/JHEP09(2020)178",
    journal = "JHEP",
    volume = "09",
    pages = "178",
    year = "2020"
}

@article{Rashkovetskyi:2021rwg,
    author = "Rashkovetskyi, Michael and Mu{\~n}oz, Julian B. and Eisenstein, Daniel J. and Dvorkin, Cora",
    title = "{Small-scale clumping at recombination and the Hubble tension}",
    eprint = "2108.02747",
    archivePrefix = "arXiv",
    primaryClass = "astro-ph.CO",
    doi = "10.1103/PhysRevD.104.103517",
    journal = "Phys. Rev. D",
    volume = "104",
    number = "10",
    pages = "103517",
    year = "2021"
}

@article{AtacamaCosmologyTelescope:2025blo,
    author = "Louis, Thibaut and others",
    collaboration = "Atacama Cosmology Telescope",
    title = "{The Atacama Cosmology Telescope: DR6 power spectra, likelihoods and {\ensuremath{\Lambda}}CDM parameters}",
    eprint = "2503.14452",
    archivePrefix = "arXiv",
    primaryClass = "astro-ph.CO",
    reportNumber = "FERMILAB-PUB-25-0071-PPD",
    doi = "10.1088/1475-7516/2025/11/062",
    journal = "JCAP",
    volume = "11",
    pages = "062",
    year = "2025"
}

@article{SPT-3G:2025bzu,
    author = "Camphuis, E. and others",
    collaboration = "SPT-3G",
    title = "{SPT-3G D1: CMB temperature and polarization power spectra and cosmology from 2019 and 2020 observations of the SPT-3G main field}",
    eprint = "2506.20707",
    archivePrefix = "arXiv",
    primaryClass = "astro-ph.CO",
    reportNumber = "FERMILAB-PUB-25-0144-PPD",
    doi = "10.1103/7wt3-9v2y",
    journal = "Phys. Rev. D",
    volume = "113",
    number = "8",
    pages = "083504",
    year = "2026"
}

@article{LiteBIRD:2023zmo,
    author = "Campeti, P. and others",
    collaboration = "LiteBIRD",
    title = "{LiteBIRD science goals and forecasts. A case study of the origin of primordial gravitational waves using large-scale CMB polarization}",
    eprint = "2312.00717",
    archivePrefix = "arXiv",
    primaryClass = "astro-ph.CO",
    doi = "10.1088/1475-7516/2024/06/008",
    journal = "JCAP",
    volume = "06",
    pages = "008",
    year = "2024"
}

@article{Balkenhol:2024sbv,
    author = "Balkenhol, L. and Trendafilova, C. and Benabed, K. and Galli, S.",
    title = "{candl: cosmic microwave background analysis with a differentiable likelihood}",
    eprint = "2401.13433",
    archivePrefix = "arXiv",
    primaryClass = "astro-ph.CO",
    doi = "10.1051/0004-6361/202449432",
    journal = "Astron. Astrophys.",
    volume = "686",
    pages = "A10",
    year = "2024"
}

@article{SPT:2019nip,
    author = "Sayre, J. T. and others",
    collaboration = "SPT",
    title = "{Measurements of B-mode Polarization of the Cosmic Microwave Background from 500 Square Degrees of SPTpol Data}",
    eprint = "1910.05748",
    archivePrefix = "arXiv",
    primaryClass = "astro-ph.CO",
    reportNumber = "FERMILAB-PUB-19-536-AE-PPD",
    doi = "10.1103/PhysRevD.101.122003",
    journal = "Phys. Rev. D",
    volume = "101",
    number = "12",
    pages = "122003",
    year = "2020"
}

@article{Balkenhol:2025xas,
    author = "Balkenhol, L. and Coerver, A. and Reichardt, C. L. and Zebrowski, J. A.",
    title = "{Probing Anisotropic Cosmic Birefringence with Foreground-Marginalised SPT B-mode Likelihoods}",
    eprint = "2510.07928",
    archivePrefix = "arXiv",
    primaryClass = "astro-ph.CO",
    reportNumber = "FERMILAB-PUB-25-0764-V",
    doi = "10.33232/001c.147459",
    journal = "Open J. Astrophys.",
    volume = "8",
    pages = "147459",
    year = "2025"
}

@article{ACT:2023dou,
    author = "Qu, Frank J. and others",
    collaboration = "ACT",
    title = "{The Atacama Cosmology Telescope: A Measurement of the DR6 CMB Lensing Power Spectrum and Its Implications for Structure Growth}",
    eprint = "2304.05202",
    archivePrefix = "arXiv",
    primaryClass = "astro-ph.CO",
    reportNumber = "FERMILAB-PUB-23-237-PPD, FERMILAB-PUB-23-237-PPD",
    doi = "10.3847/1538-4357/acfe06",
    journal = "Astrophys. J.",
    volume = "962",
    number = "2",
    pages = "112",
    year = "2024"
}

@article{ACT:2023ubw,
    author = "MacCrann, Niall and others",
    collaboration = "ACT",
    title = "{The Atacama Cosmology Telescope: Mitigating the Impact of Extragalactic Foregrounds for the DR6 Cosmic Microwave Background Lensing Analysis}",
    eprint = "2304.05196",
    archivePrefix = "arXiv",
    primaryClass = "astro-ph.CO",
    reportNumber = "FERMILAB-PUB-23-236-PPD",
    doi = "10.3847/1538-4357/ad2610",
    journal = "Astrophys. J.",
    volume = "966",
    number = "1",
    pages = "138",
    year = "2024"
}

@article{SPT-3G:2024atg,
    author = "Ge, F. and others",
    collaboration = "SPT-3G",
    title = "{Cosmology from CMB lensing and delensed EE power spectra using 2019{\textendash}2020 SPT-3G polarization data}",
    eprint = "2411.06000",
    archivePrefix = "arXiv",
    primaryClass = "astro-ph.CO",
    reportNumber = "FERMILAB-PUB-24-0840-PPD",
    doi = "10.1103/PhysRevD.111.083534",
    journal = "Phys. Rev. D",
    volume = "111",
    number = "8",
    pages = "083534",
    year = "2025"
}

@article{ACT:2023kun,
    author = "Madhavacheril, Mathew S. and others",
    collaboration = "ACT",
    title = "{The Atacama Cosmology Telescope: DR6 Gravitational Lensing Map and Cosmological Parameters}",
    eprint = "2304.05203",
    archivePrefix = "arXiv",
    primaryClass = "astro-ph.CO",
    reportNumber = "FERMILAB-PUB-23-206-PPD",
    doi = "10.3847/1538-4357/acff5f",
    journal = "Astrophys. J.",
    volume = "962",
    number = "2",
    pages = "113",
    year = "2024"
}

@article{Caloni:2026dyu,
    author = "Caloni, Luca and Ferreira, Ricardo Z. and Sousa, Lara and Winckler, Clara",
    title = "{Cosmic strings and domain walls: the impact of CMB $B$-mode data}",
    eprint = "2602.20050",
    archivePrefix = "arXiv",
    primaryClass = "astro-ph.CO",
    month = "2",
    year = "2026"
}

@article{SPT-3G:2025vtb,
    author = "Zebrowski, J. A. and others",
    collaboration = "SPT-3G",
    title = "{SPT-3G D1: Constraints on inflationary gravitational waves with two years of SPT-3G data}",
    eprint = "2505.02827",
    archivePrefix = "arXiv",
    primaryClass = "astro-ph.CO",
    reportNumber = "FERMILAB-PUB-25-0319-PPD",
    doi = "10.1103/33wm-c6pt",
    journal = "Phys. Rev. D",
    volume = "112",
    number = "12",
    pages = "123520",
    year = "2025"
}

@article{Eimer:2023esh,
    author = "Eimer, Joseph R. and others",
    title = "{CLASS Angular Power Spectra and Map-component Analysis for 40 GHz Observations through 2022}",
    eprint = "2309.00675",
    archivePrefix = "arXiv",
    primaryClass = "astro-ph.CO",
    doi = "10.3847/1538-4357/ad1abf",
    journal = "Astrophys. J.",
    volume = "963",
    number = "2",
    pages = "92",
    year = "2024"
}

@article{Micheli:2024hfe,
    author = "Micheli, Silvia and others",
    title = "{Systematic effects induced by half-wave plate differential optical load and TES nonlinearity for LiteBIRD}",
    eprint = "2407.15294",
    archivePrefix = "arXiv",
    primaryClass = "astro-ph.IM",
    doi = "10.1117/12.3018553",
    journal = "Proc. SPIE Int. Soc. Opt. Eng.",
    volume = "13102",
    pages = "131022R",
    year = "2024"
}

@article{Giardiello:2021uxq,
    author = "Giardiello, Serena and others",
    title = "{Detailed study of HWP non-idealities and their impact on future measurements of CMB polarization anisotropies from space}",
    eprint = "2106.08031",
    archivePrefix = "arXiv",
    primaryClass = "astro-ph.CO",
    doi = "10.1051/0004-6361/202141619",
    journal = "Astron. Astrophys.",
    volume = "658",
    pages = "A15",
    year = "2022",
    note = "[Erratum: Astron.Astrophys. 671, C1 (2023)]"
}

@article{Kostelecky:2002hh,
    author = "Kosteleck{\'y}, V. Alan and Mewes, Matthew",
    title = "{Signals for Lorentz violation in electrodynamics}",
    eprint = "hep-ph/0205211",
    archivePrefix = "arXiv",
    reportNumber = "IUHET-449",
    doi = "10.1103/PhysRevD.66.056005",
    journal = "Phys. Rev. D",
    volume = "66",
    pages = "056005",
    year = "2002"
}

@article{Galloni:2024lre,
    author = "Galloni, Giacomo and Henrot-Versill{\'e}, Sophie and Tristram, Matthieu",
    title = "{Robust constraints on tensor perturbations from cosmological data: A comparative analysis from Bayesian and frequentist perspectives}",
    eprint = "2405.04455",
    archivePrefix = "arXiv",
    primaryClass = "astro-ph.CO",
    doi = "10.1103/PhysRevD.110.063511",
    journal = "Phys. Rev. D",
    volume = "110",
    number = "6",
    pages = "063511",
    year = "2024"
}

@article{Tristram:2020wbi,
    author = "Tristram, M. and others",
    title = "{Planck constraints on the tensor-to-scalar ratio}",
    eprint = "2010.01139",
    archivePrefix = "arXiv",
    primaryClass = "astro-ph.CO",
    doi = "10.1051/0004-6361/202039585",
    journal = "Astron. Astrophys.",
    volume = "647",
    pages = "A128",
    year = "2021"
}

@article{Tristram:2021tvh,
    author = "Tristram, M. and others",
    title = "{Improved limits on the tensor-to-scalar ratio using BICEP and Planck data}",
    eprint = "2112.07961",
    archivePrefix = "arXiv",
    primaryClass = "astro-ph.CO",
    doi = "10.1103/PhysRevD.105.083524",
    journal = "Phys. Rev. D",
    volume = "105",
    number = "8",
    pages = "083524",
    year = "2022"
}

@article{Essinger-Hileman:2026tgi,
    author = "Essinger-Hileman, Thomas and others",
    title = "{Improved Constraints on Cosmic Microwave Background Circular Polarization with CLASS}",
    eprint = "2609.40125",
    archivePrefix = "arXiv",
    primaryClass = "astro-ph.CO",
    month = "9",
    year = "2026"
}

@article{Kostelecky:2013rv,
    author = "Kosteleck{\'y}, V. Alan and Mewes, Matthew",
    title = "{Constraints on relativity violations from gamma-ray bursts}",
    eprint = "1301.5367",
    archivePrefix = "arXiv",
    primaryClass = "astro-ph.HE",
    reportNumber = "IUHET-572, JANUARY-2013",
    doi = "10.1103/PhysRevLett.110.201601",
    journal = "Phys. Rev. Lett.",
    volume = "110",
    number = "20",
    pages = "201601",
    year = "2013"
}

@article{Kostelecky:2006ta,
    author = "Kosteleck{\'y}, V. Alan and Mewes, Matthew",
    title = "{Sensitive polarimetric search for relativity violations in gamma-ray bursts}",
    eprint = "hep-ph/0607084",
    archivePrefix = "arXiv",
    reportNumber = "IUHET-492",
    doi = "10.1103/PhysRevLett.97.140401",
    journal = "Phys. Rev. Lett.",
    volume = "97",
    pages = "140401",
    year = "2006"
}

@article{Diego-Palazuelos:2025dmh,
    author = "Diego-Palazuelos, Patricia and Komatsu, Eiichiro",
    title = "{Cosmic birefringence from the Atacama Cosmology Telescope Data Release 6}",
    eprint = "2509.13654",
    archivePrefix = "arXiv",
    primaryClass = "astro-ph.CO",
    doi = "10.1103/pbc3-t52s",
    journal = "Phys. Rev. D",
    volume = "113",
    number = "10",
    pages = "L101302",
    year = "2026"
}

@article{SimonsObservatory:2025avm,
    author = "Abril-Cabezas, I. and others",
    collaboration = "Simons Observatory",
    title = "{The Simons Observatory: forecasted constraints on primordial gravitational waves with the expanded array of Small Aperture Telescopes}",
    eprint = "2512.15833",
    archivePrefix = "arXiv",
    primaryClass = "astro-ph.CO",
    reportNumber = "FERMILAB-PUB-25-0947-PPD",
    doi = "10.1088/1475-7516/2026/04/051",
    journal = "JCAP",
    volume = "04",
    pages = "051",
    year = "2026"
}

@article{LiteBIRD:2026qdz,
    author = "Raffuzzi, N. and others",
    collaboration = "LiteBIRD",
    title = "{Mitigating half-wave plate systematics at the map-making level: calibration requirements for LiteBIRD}",
    eprint = "2602.02410",
    archivePrefix = "arXiv",
    primaryClass = "astro-ph.CO",
    doi = "10.1088/1475-7516/2026/08/053",
    journal = "JCAP",
    volume = "08",
    pages = "053",
    year = "2026"
}

@article{Komatsu034005,
author = {Kunimoto Komatsu and Hirokazu Ishino and Nobuhiko Katayama and Tomotake Matsumura and Yuki Sakurai},
title = {{Design of a frequency-independent optic axis Pancharatnam-based achromatic half-wave plate}},
volume = {7},
journal = {Journal of Astronomical Telescopes, Instruments, and Systems},
number = {3},
publisher = {SPIE},
pages = {034005},
year = {2021},
doi = {10.1117/1.JATIS.7.3.034005},
URL = {https://doi.org/10.1117/1.JATIS.7.3.034005}
}

@article{Pisano:2020qxq,
    author = "Pisano, G. and others",
    title = "{Development and application of metamaterial-based half-wave plates for the NIKA and NIKA2 polarimeters}",
    eprint = "2006.12081",
    archivePrefix = "arXiv",
    primaryClass = "astro-ph.IM",
    doi = "10.1051/0004-6361/202038643",
    journal = "Astron. Astrophys.",
    volume = "658",
    pages = "A24",
    year = "2022"
}

@article{Kushwaha:2025yja,
    author = "Kushwaha, Ashu and Jain, Rajeev Kumar",
    title = "{Probing axion photon conversion via circular polarization imprints in the CMB V-mode observations}",
    eprint = "2510.22592",
    archivePrefix = "arXiv",
    primaryClass = "astro-ph.CO",
    doi = "10.1103/32gt-4199",
    journal = "Phys. Rev. D",
    volume = "113",
    number = "12",
    pages = "123511",
    year = "2026"
}

@article{Hoseinpour:2020hic,
    author = "Hoseinpour, Ahmad and Zarei, Moslem and Orlando, Giorgio and Bartolo, Nicola and Matarrese, Sabino",
    title = "{CMB $V$ modes from photon-photon forward scattering revisited}",
    eprint = "2006.14418",
    archivePrefix = "arXiv",
    primaryClass = "hep-ph",
    doi = "10.1103/PhysRevD.102.063501",
    journal = "Phys. Rev. D",
    volume = "102",
    number = "6",
    pages = "063501",
    year = "2020"
}

@article{SimonsObservatory:2025wwn,
    author = "Abitbol, M. and others",
    collaboration = "Simons Observatory",
    title = "{The Simons Observatory: science goals and forecasts for the enhanced Large Aperture Telescope}",
    eprint = "2503.00636",
    archivePrefix = "arXiv",
    primaryClass = "astro-ph.IM",
    reportNumber = "FERMILAB-PUB-25-0188-PPD",
    doi = "10.1088/1475-7516/2025/08/034",
    journal = "JCAP",
    volume = "08",
    pages = "034",
    year = "2025"
}

@article{Kasai:2026qty,
    author = "Kasai, Kentaro and Obata, Ippei",
    title = "{Cosmic Birefringence from Large Lepton Asymmetry}",
    eprint = "2610.09748",
    archivePrefix = "arXiv",
    primaryClass = "astro-ph.CO",
    reportNumber = "KEK-TH-2879, KEK-Cosmo-0436, RUP-26-28",
    month = "10",
    year = "2026"
}

@article{Zhou:2023aqz,
    author = "Zhou, Ren-Peng and Huang, Da and Geng, Chao-Qiang",
    title = "{Cosmic birefringence from neutrino and dark matter asymmetries}",
    eprint = "2302.11140",
    archivePrefix = "arXiv",
    primaryClass = "astro-ph.CO",
    doi = "10.1088/1475-7516/2023/07/053",
    journal = "JCAP",
    volume = "07",
    pages = "053",
    year = "2023"
}

@article{Murai:2022zur,
    author = "Murai, Kai and Naokawa, Fumihiro and Namikawa, Toshiya and Komatsu, Eiichiro",
    title = "{Isotropic cosmic birefringence from early dark energy}",
    eprint = "2209.07804",
    archivePrefix = "arXiv",
    primaryClass = "astro-ph.CO",
    reportNumber = "RESCEU-16/22",
    doi = "10.1103/PhysRevD.107.L041302",
    journal = "Phys. Rev. D",
    volume = "107",
    number = "4",
    pages = "L041302",
    year = "2023"
}

@article{Kitajima:2022jzz,
    author = "Kitajima, Naoya and Kozai, Fumiaki and Takahashi, Fuminobu and Yin, Wen",
    title = "{Power spectrum of domain-wall network, and its implications for isotropic and anisotropic cosmic birefringence}",
    eprint = "2205.05083",
    archivePrefix = "arXiv",
    primaryClass = "astro-ph.CO",
    reportNumber = "TU-1155",
    doi = "10.1088/1475-7516/2022/10/043",
    journal = "JCAP",
    volume = "10",
    pages = "043",
    year = "2022"
}

@article{Obata:2021nql,
    author = "Obata, Ippei",
    title = "{Implications of the cosmic birefringence measurement for the axion dark matter search}",
    eprint = "2108.02150",
    archivePrefix = "arXiv",
    primaryClass = "astro-ph.CO",
    doi = "10.1088/1475-7516/2022/09/062",
    journal = "JCAP",
    volume = "09",
    pages = "062",
    year = "2022"
}

@article{Takahashi:2020tqv,
    author = "Takahashi, Fuminobu and Yin, Wen",
    title = "{Kilobyte Cosmic Birefringence from ALP Domain Walls}",
    eprint = "2012.11576",
    archivePrefix = "arXiv",
    primaryClass = "hep-ph",
    reportNumber = "TU-1115, IPMU20-0130",
    doi = "10.1088/1475-7516/2021/04/007",
    journal = "JCAP",
    volume = "04",
    pages = "007",
    year = "2021"
}

@article{Fujita:2020ecn,
    author = "Fujita, Tomohiro and Murai, Kai and Nakatsuka, Hiromasa and Tsujikawa, Shinji",
    title = "{Detection of isotropic cosmic birefringence and its implications for axionlike particles including dark energy}",
    eprint = "2011.11894",
    archivePrefix = "arXiv",
    primaryClass = "astro-ph.CO",
    doi = "10.1103/PhysRevD.103.043509",
    journal = "Phys. Rev. D",
    volume = "103",
    number = "4",
    pages = "043509",
    year = "2021"
}

@article{Geng:2007va,
    author = "Geng, C. Q. and Ho, S. H. and Ng, J. N.",
    title = "{Neutrino number asymmetry and cosmological birefringence}",
    eprint = "0706.0080",
    archivePrefix = "arXiv",
    primaryClass = "astro-ph",
    doi = "10.1088/1475-7516/2007/09/010",
    journal = "JCAP",
    volume = "09",
    pages = "010",
    year = "2007"
}

@article{Gasparotto:2023psh,
    author = "Gasparotto, Silvia and Sfakianakis, Evangelos I.",
    title = "{Cosmic birefringence from the Axiverse}",
    eprint = "2306.16355",
    archivePrefix = "arXiv",
    primaryClass = "astro-ph.CO",
    doi = "10.1088/1475-7516/2023/11/017",
    journal = "JCAP",
    volume = "11",
    pages = "017",
    year = "2023"
}

@article{Gasparotto:2022uqo,
    author = "Gasparotto, Silvia and Obata, Ippei",
    title = "{Cosmic birefringence from monodromic axion dark energy}",
    eprint = "2203.09409",
    archivePrefix = "arXiv",
    primaryClass = "astro-ph.CO",
    doi = "10.1088/1475-7516/2022/08/025",
    journal = "JCAP",
    volume = "08",
    number = "08",
    pages = "025",
    year = "2022"
}

@article{POLARBEAR:2026zav,
    author = "Arnold, Kam and others",
    collaboration = "POLARBEAR",
    title = "{Millimeter-Wavelength Constraints on Circular Polarization of the Cosmic Microwave Background at Arcminute Scales with the Simons Array}",
    eprint = "2611.13345",
    archivePrefix = "arXiv",
    primaryClass = "astro-ph.CO",
    month = "11",
    year = "2026"
}

\end{document}